# The affine and Euclidean normalizers of the subperiodic groups

**Brian Kevin VanLeeuwen[a]\*, Pedro Valentín De Jesús[a], Daniel B. Litvin[b], Venkatraman Gopalan[a]**
[a]Department of Materials Science and Engineering, The Pennsylvania State University, University Park, Pennsylvania, 16803, USA, and [b]Department of Physics, The Eberly College of Science, The Pennsylvania State University, Penn State Berks, P.O. Box 7009, Reading, Pennsylvania, 19610, USA

Correspondence email: Brian.VanLeeuwen@gmail.com

**Synopsis**

The affine and Euclidean normalizers of the subperiodic groups are derived and listed.

**Abstract**

The affine and Euclidean normalizers of the subperiodic groups, the Frieze groups, the rod groups, and the layer groups, are derived and listed. For the layer groups, the special metrics used for plane group Euclidean normalizers have been considered.

**1. Introduction**

Informally, a normalizer of a symmetry group can be thought of as the symmetry group of that symmetry group, *i.e.* it is the set of transformations that map the group into itself. For instance, with the point group 222 which consists of three perpendicular 2-fold rotation axes, a 90 degree rotation about one of the 2-fold axes would not change the group. Similarly, mirroring across planes perpendicular to one of the 2-fold axes would not change 222. Taking all such operations together, we would find that the group $m\bar{3}m$ is formed. That is to say that $m\bar{3}m$ is the orthogonal normalizer group of 222.

Formally, a given group $S$ and a subgroup $G$ of $S$, the normalizer group of $G$ in $S$ is the subgroup of $S$ composed of every element $s$ in $S$ for which $G = sGs^{-1}$ [1] (Boisen et al., 1990; Opechowski, 1986; Hahn, 2006; Koch et al. 2006). Note that it is not necessary for every element $g$ in $G$ to satisfy $g = sgs^{-1}$, just $G$ as a whole. The normalizer group of $G$ in $S$ will be denoted $N_S(G)$.

$$N_S(G) = \{s \in S : G = sGs^{-1}\}$$

---

[1] Note that expressions such as $aSa^{-1}$ where $S$ is a set or a group and $a$ is a transformation should be interpreted as applying $asa^{-1}$ to every element $s$ in $S$, *i.e.* $aSa^{-1} = \{asa^{-1} : s \in S\}$.

The *affine normalizer group of a space group* is the normalizer group of that space group in the group of affine transformations. Likewise, the *Euclidean normalizer group of a space group* is the normalizer group of that space group in the group of Euclidean motions. The group of affine transformations will be denoted $A$ and the group of Euclidean motions will be denoted $E$. When discussing the rod and layer groups, $A$ and $E$ will refer specifically to the affine and Euclidean groups of three-dimensional Euclidean space. When discussing the Frieze groups, they will refer to the affine and Euclidean group of two-dimensional Euclidean space. To avoid excessive notation, context will distinguish the two and three dimensional cases.

It has been suggested that our listing of the double antisymmetry space group properties and diagrams (Huang et al. 2014) should include the double antisymmetry subperiodic groups similar to Litvin's listing of the 1-, 2-, and 3-dimensional magnetic space and subperiodic groups (Litvin 2013). This was not possible because the types of double antisymmetry subperiodic groups are not known and the method of evaluating proper affine equivalence discussed by VanLeeuwen et al. (2014) requires using the affine normalizer groups. Thus, the motivation for the present work, in addition to the value of the normalizer groups themselves, was to make it possible to use this method for the subperiodic antisymmetry groups. Although listing the Euclidean normalizers is not necessary for this purpose, it is a result that may be of some interest to others and we have therefore extended the results to cover this as well.

## 2. Affine normalizers of a translational subgroup

In space groups and subperiodic groups of every dimension, every element is composed of a distance-preserving linear transformation and a translation, such a transformation is called a *motion* or a *Euclidean motion*. A motion is a special type of the more general *affine transformations*. Affine transformations are composed of linear transformations and translations. Affine transformations, unlike motions, do not require the linear part to be distance preserving; hence $E \subset A$. In the present work, affine transformations (and motions as a special case) will be denoted in Seitz notation as $\{a|\vec{t}\}$ where $a$ is the linear part and $\vec{t}$ is the translation vector.

The product of affine transformations $\{a_1|\vec{t}_1\}$ and $\{a_2|\vec{t}_2\}$ can be written in Seitz notation as $\{a_1|\vec{t}_1\}\{a_2|\vec{t}_2\} = \{a_1 a_2 | a_1 \vec{t}_2 + \vec{t}_1\}$. The inverse of an affine transformation $\{a|\vec{t}\}$ must therefore be $\{a|\vec{t}\}^{-1} = \{a^{-1}| -a^{-1}\vec{t}\}$. Similarity transformations can be performed using these Seitz notation expressions.

The similarity transformation of a pure translation (an affine transformation with identity for the linear part) by an affine transformation is also a pure translation. This is easy to prove by showing that $\{a|\vec{t}\}\{1|\vec{t}'\}\{a^{-1}|-a^{-1}\vec{t}\}$ simplifies to $\{1|a\vec{t}'\}$. Since the similarity transformation of a pure translation by an affine transformation is also a pure translation, an element of the affine normalizer of a symmetry group is also an element of the affine normalizer of the translational subgroup [2], i.e., $G = nGn^{-1}$ implies $\Lambda = n\Lambda n^{-1}$ for all for $G \subset E$. Therefore $N_A(G) \subset N_A(\Lambda)$.

Let $G$ be a group of Euclidean motions that transform $d$-dimensional Euclidean space and $\Lambda$ be the translational subgroup, if the translation vectors of $\Lambda$ span the $d$-dimensional space and $\Lambda \cong \mathbb{Z}^d$, then it is a $d$-dimensional space group, if they span a zero-dimensional subspace, it is an $d$-dimensional point group, and if they span a $p$-dimensional subspace and $\Lambda \cong \mathbb{Z}^p$ where $d > p > 0$, it is a $d$-dimensional subperiodic group with $p$ dimensions of periodicity. Figure 1 shows how subperiodic groups fill the gap between point and space groups. We will consider a faithful augmented matrix representation of the group $G$ on a basis of $p$ primitive lattice vectors and $d - p$ arbitrary vectors that form an orthogonal basis for the orthogonal complement of the aforementioned $p$-dimensional subspace. This representation of $G$ on this basis and the abstract group itself will not be distinguished. Similarly, the faithful representations of $A$ and $E$ on this basis will not be distinguished from $A$ and $E$. This representation of $G$ was achieved by a basis transformation from the standard augmented matrix representation in the standard setting using the appropriate centering matrix. The standard augmented matrices we used were downloaded from the Bilbao Crystallographic Server (Aroyo et al. 2006a; Aroyo et al. 2006b; Aroyo et al. 2011).

Since $N_A(G) \subset N_A(\Lambda)$, to solve for $N_A(G)$ we will start by deriving the form of all elements of $N_A(\Lambda)$. On this generalized primitive basis, the augmented matrix of a general affine transformation would be represented by a $d$-by-$d$ invertible matrix of real numbers for the linear part and $d$ real number components for the translation vector. This can be represented as a block matrix $\left\{\begin{matrix} n & r_\| \\ c & r_\perp \end{matrix} \middle| \begin{matrix} \vec{x}_\| \\ \vec{x}_\perp \end{matrix}\right\}$ where $n$ is a $p$-by-$p$ submatrix, $r_\|$ is a $p$-by-$(d-p)$ submatrix, $r_\perp$ is $(d-p)$-by-$(d-p)$ submatrix, $\vec{x}_\|$ is a $p$-by-1 submatrix (i.e. a $p$-dimensional "column vector") and $\vec{x}_\perp$ is a $(d-p)$-by-1 submatrix (i.e. a $(d-p)$-dimensional "column vector"). If this affine transformation is an element of $N_A(\Lambda)$, it should map elements of $\Lambda$ into elements of $\Lambda$. On this basis, elements of $\Lambda$ are $\left\{1 \middle| \begin{matrix} \vec{t} \\ 0 \end{matrix}\right\}$ where 1 is a $d$-by-$d$ identity matrix, $\vec{t}$ is a $p$-by-1 submatrix of integers and 0 is a $(d-p)$-by-1 zero submatrix. Conjugation of $\left\{1 \middle| \begin{matrix} \vec{t} \\ 0 \end{matrix}\right\}$ by $\left\{\begin{matrix} n & r_\| \\ c & r_\perp \end{matrix} \middle| \begin{matrix} \vec{x}_\| \\ \vec{x}_\perp \end{matrix}\right\}$ yields $\left\{1 \middle| \begin{matrix} n\vec{t} \\ c\vec{t} \end{matrix}\right\}$. For $\left\{1 \middle| \begin{matrix} n\vec{t} \\ c\vec{t} \end{matrix}\right\}$ to be an element of $\Lambda$, the components of $n\vec{t}$ must be integers

---

[2] The translational subgroup of a group $G \subset E$ will be denoted $\Lambda$ and defined by $\Lambda \equiv \{\{1|\vec{t}\} \in G\}$.

and $c\vec{t} = \vec{0}$. Thus for $\left\{\begin{matrix} n & r_\parallel \\ c & r_\perp \end{matrix} \middle| \begin{matrix} \vec{x}_\parallel \\ \vec{x}_\perp \end{matrix}\right\}$ to be an element of $N_A(\Lambda)$, $c$ must be the zero matrix and $n$ must be a matrix of integers. Furthermore, $|n| = \pm 1$ because the volume of the subperiodic unit cell must be conserved. Given that $c$ must be the zero matrix, we can also find that, in order to guarantee invertibility, $|r_\perp|$ must be non-zero, because $\begin{vmatrix} n & r_\parallel \\ 0 & r_\perp \end{vmatrix} = |n||r_\perp|$.

Thus, the elements of $N_A(\Lambda)$ are simply the augmented matrices with the following form:

$$\left\{\begin{matrix} n & r_\parallel \\ 0 & r_\perp \end{matrix} \middle| \vec{x}\right\}$$

where $n$ is a $p$-by-$p$ submatrix of integers with $|n| = \pm 1$, $r_\parallel$ is a $p$-by-$(d-p)$ submatrix of real numbers, $r_\perp$ is $(d-p)$-by-$(d-p)$ submatrix of real numbers with $|r_\perp| \neq 0$, and $\vec{x}$ is the translational vector whose coordinates are real numbers.

For the Frieze groups, which are two dimensional subperiodic groups with one periodic dimension ($d = 2, p = 1$):

$$N_A(\Lambda) = \left\{\left\{\begin{matrix} \pm 1 & r_{12} \\ 0 & r_{22} \end{matrix} \middle| \begin{matrix} x_1 \\ x_2 \end{matrix}\right\} : r_{22} \neq 0 \wedge r_{ij}, x_i \in \mathbb{R}\right\}$$

For the rod groups, which are three dimensional subperiodic groups with one periodic dimension ($d = 3, p = 1$) with the translations along [100] rather than the rod group convention of [001] :

$$N_A(\Lambda) = \left\{\left\{\begin{matrix} \pm 1 & r_{12} & r_{13} \\ 0 & r_{22} & r_{23} \\ 0 & r_{32} & r_{33} \end{matrix} \middle| \begin{matrix} x_1 \\ x_2 \\ x_3 \end{matrix}\right\} : \begin{vmatrix} r_{22} & r_{23} \\ r_{32} & r_{33} \end{vmatrix} \neq 0 \wedge r_{ij}, x_i \in \mathbb{R}\right\}$$

Rearranged for translations along [001]:

$$N_A(\Lambda) = \left\{\left\{\begin{matrix} r_{11} & r_{12} & 0 \\ r_{21} & r_{22} & 0 \\ r_{31} & r_{32} & \pm 1 \end{matrix} \middle| \begin{matrix} x_1 \\ x_2 \\ x_3 \end{matrix}\right\} : \begin{vmatrix} r_{11} & r_{12} \\ r_{21} & r_{22} \end{vmatrix} \neq 0 \wedge r_{ij}, x_i \in \mathbb{R}\right\}$$

For the layer groups, which are three dimensional subperiodic groups with two periodic dimensions ($d = 3, p = 2$):

$$N_A(\Lambda) = \left\{\left\{\begin{matrix} n_{11} & n_{12} & r_{13} \\ n_{21} & n_{22} & r_{23} \\ 0 & 0 & r_{33} \end{matrix} \middle| \begin{matrix} x_1 \\ x_2 \\ x_3 \end{matrix}\right\} : \begin{vmatrix} n_{11} & n_{12} \\ n_{21} & n_{22} \end{vmatrix} = \pm 1 \wedge r_{33} \neq 0 \wedge r_{ij}, x_i \in \mathbb{R} \wedge n_{ij} \in \mathbb{Z}\right\}$$

A different approach to deriving normalizers of crystallographic groups was discussed by Boisen et al. (1990). One of the main differences is that Boisen et al. start from the point group normalizers, noting that the linear part of an element of a space group normalizer must be an element of the normalizer of its point group. This would not be a significant benefit for deriving the subperiodic group normalizers because it would only constrain the $p$-by-$p$ submatrix of integers.

## 3. Affine normalizers of subperiodic groups

The problem of computing the normalizer of a group $G \subset E$ will be simplified by considering the quotient of $G$ with respect to its translational subgroup $\Lambda$. $G$ is factored by $\Lambda$ into a set of cosets:

$$G/\Lambda = \{g_1\Lambda, g_2\Lambda, \ldots, g_i\Lambda\}$$

where $g_1$ through $g_i$ is an indexed set of coset representatives (the identity motion will be selected for $g_1$). Since we are only considering crystallographic subperiodic groups, $i$ is finite and equal to the size of the point group of $G$, e.g. the point group of $pmm2$ is $mm2$ which has four elements thus $i=4$ for $G = pmm2$. Let $s$ denote an element of the affine normalizer of $G$. Since $G = sGs^{-1}$, we may write that

$$G/\Lambda = \{sg_1\Lambda s^{-1}, sg_2\Lambda s^{-1}, \ldots, sg_i\Lambda s^{-1}\}$$

As discussed in Section 2, $G = sGs^{-1}$ implies $\Lambda = s\Lambda s^{-1}$. We can therefore substitute $s^{-1}\Lambda s$ for $\Lambda$:

$$G/\Lambda = \{sg_1(s^{-1}\Lambda s)s^{-1}, sg_2(s^{-1}\Lambda s)s^{-1}, \ldots, sg_i(s^{-1}\Lambda s)s^{-1}\}$$

which simplifies to

$$G/\Lambda = \{sg_1s^{-1}\Lambda, sg_2s^{-1}\Lambda, \ldots, sg_is^{-1}\Lambda\}$$

Thus, if $s$ is an element of the normalizer of $G$, $sg_1s^{-1}\Lambda$ through $sg_is^{-1}\Lambda$ is just a permutation of $g_1\Lambda$ through $g_i\Lambda$ that preserves the group law of $G/\Lambda$, i.e. an automorphism of $G/\Lambda$. Furthermore, this is only true for affine transformations when $s$ is an element of the affine normalizer of $G$ because $G$ is the union of the cosets of $G/\Lambda$. Therefore, an affine transformation $s$ is an element of the affine normalizer of $G$ if and only if

$$(sg_1s^{-1}\Lambda, sg_2s^{-1}\Lambda, \ldots, sg_is^{-1}\Lambda) = \sigma(g_1\Lambda, g_2\Lambda, \ldots, g_i\Lambda)$$

for some $\sigma$ in Aut($G/\Lambda$). This can be further refined by the realization that $g_j\Lambda$ is conjugate to $sg_js^{-1}\Lambda$ in $N_A(\Lambda)$ because $N_A(G) \subset N_A(\Lambda)$ [3]. We can therefore limit $\sigma$ to only those automorphisms of $G/\Lambda$ which respect this condition. Furthermore, we need only consider a generating set of $G/\Lambda$, provided that the generating set is closed under conjugacy in $N_A(\Lambda)$. It is also possible to avoid the difficulty of computing inverses by rearranging the previous equation to the following:

$$(sg_1\Lambda, sg_2\Lambda, \ldots, sg_i\Lambda) = \sigma(g_1s\Lambda, g_2s\Lambda, \ldots, g_is\Lambda)$$

---

[3] Note that this is not quite the same as $g_j$ is conjugate to $sg_js^{-1}$ in $N_A(\Lambda)$ because the elements of $g_j\Lambda$ can fall into two separate conjugacy classes of $N_A(\Lambda)$ if some of its elements have no glide components and others have non-primitive glides. In three or fewer dimensions, the glide vector of a crystallographic motion $\{a|\vec{t}\}$ can be calculated by $\vec{t}_{glide} = \frac{\sum_{i=0}^{11} a^i}{12}\vec{t}$ (a consequence of the LCM of the orders allowed by the crystallographic restriction theorem). If a primitive basis is used, then the conjugacy of $\{a|\vec{t}\}\Lambda$ and $s\{a|\vec{t}\}s^{-1}\Lambda$ on $N_A(\Lambda)$ can be checked by comparing the elements of each where the translation components are greater than or equal to zero and less than one.

Determining conjugacy on $N_A(\Lambda)$ is fairly straightforward in Mathematica by setting up the appropriate equations and checking for solutions using the symbolic solver, but may be difficult without comparable software. A much simpler condition is to only consider permutations that exchange elements with identical invariant characteristics, *i.e.* class functions on $N_A(\Lambda)$ or $A$ (*e.g.* determinants, traces, eigenvalues, *etc.*). It is also possible to use canonical forms, if non-primitive glides and possible reorderings of blocks are carefully accounted for.

To summarize, for each subperiodic group $G$, $N_A(G)$ is solved for by:

1. factoring $G$ into a finite set of cosets with respect to $\Lambda$,

2. considering how those cosets may be permuted by elements of $N_A(\Lambda)$, and then

3. solving for the elements of $N_A(\Lambda)$ that are also elements of $N_A(G)$.

### 3.1. Example: L23, *pmm*2

In this section, we will demonstrate how $N_A(G)$ can be solved for with a specific example: $G = pmm2$, the 23$^{\text{rd}}$ type of layer group. For this group,

$$G/\Lambda = \{\{1|0\}\Lambda, \{2_{001}|0\}\Lambda, \{m_{100}|0\}\Lambda, \{m_{010}|0\}\Lambda\}$$

There are 4! = 24 permutations of $G/\Lambda$, but most are not automorphisms and even fewer preserve conjugacy on $N_A(\Lambda)$. Checking the additional permutations is not necessary but will not affect the results. For this group, 24 permutations is computationally tractable but for other groups the number of permutations is too large. For instance, with $p6/mmm$, $|G/\Lambda| = 24$ and $24! \sim 6 \times 10^{23}$; clearly brute force computing this many permutations is not going to be possible!

The condition that conjugacy on $N_A(\Lambda)$ needs to be preserved means that the only permutations of $G/\Lambda$ that need to be considered are the identity permutation and the transposition of $\{m_{100}|0\}\Lambda$ and $\{m_{010}|0\}\Lambda$ (in this case, the invariance of the eigenvalues would have been enough to eliminate the other 22 permutations). These two permutations are depicted in Figure 2. Thus, the affine normalizers have to satisfy the conditions (1), (2) and either (3a) or (3b):

(1): $s\{1|0\}\Lambda = \{1|0\}s\Lambda$

(2): $s\{2_{001}|0\}\Lambda = \{2_{001}|0\}s\Lambda$

(3a): $s\{m_{100}|0\}\Lambda = \{m_{100}|0\}s\Lambda$ and $s\{m_{010}|0\}\Lambda = \{m_{010}|0\}s\Lambda$

(3b): $s\{m_{100}|0\}\Lambda = \{m_{010}|0\}s\Lambda$ and $s\{m_{010}|0\}\Lambda = \{m_{100}|0\}s\Lambda$

Putting these into explicit algebraic will only be shown for (2). Recalling that $\Lambda = \{\{1|h,k,0\} : h,k \in \mathbb{Z}\}$ for a primitive basis of a layer group, $s\{2_{001}|0\}\Lambda$ can be written explicitly with matrices as

$$\left\{ \begin{pmatrix} n_{11} & n_{12} & r_{13} & x_1 \\ n_{21} & n_{22} & r_{23} & x_2 \\ 0 & 0 & r_{33} & x_3 \end{pmatrix} \begin{pmatrix} -1 & 0 & 0 & 0 \\ 0 & -1 & 0 & 0 \\ 0 & 0 & 1 & 0 \end{pmatrix} \begin{pmatrix} 1 & 0 & 0 & h \\ 0 & 1 & 0 & k \\ 0 & 0 & 1 & 0 \end{pmatrix} : h, k \in \mathbb{Z} \right\}$$

which may be simplified to

$$\left\{ \begin{pmatrix} -n_{11} & -n_{12} & r_{13} & -h\,n_{11} - k\,n_{12} + x_1 \\ -n_{21} & -n_{22} & r_{23} & -h\,n_{21} - k\,n_{22} + x_2 \\ 0 & 0 & r_{33} & x_3 \end{pmatrix} : h, k \in \mathbb{Z} \right\}$$

Similarly $\{2_{001}|0\}s\Lambda$ may be written:

$$\left\{ \begin{pmatrix} -n_{11} & -n_{12} & -r_{13} & -h\,n_{11} - k\,n_{12} - x_1 \\ -n_{21} & -n_{22} & -r_{23} & -h\,n_{21} - k\,n_{22} - x_2 \\ 0 & 0 & r_{33} & x_3 \end{pmatrix} : h, k \in \mathbb{Z} \right\}$$

Accounting for terms that are fixed across the sets, the logical expansion over the components of $s\{2_{001}|0\}\Lambda = \{2_{001}|0\}s\Lambda$ thus simplifies to $r_{13} = -r_{13}$, $r_{23} = -r_{23}$, $\{-h\,n_{11} - k\,n_{12} + x_1 : h, k \in \mathbb{Z}\} = \{-h\,n_{11} - k\,n_{12} - x_1 : h, k \in \mathbb{Z}\}$, and $\{-h\,n_{21} - k\,n_{22} + x_2 : h, k \in \mathbb{Z}\} = \{-h\,n_{21} - k\,n_{22} - x_2 : h, k \in \mathbb{Z}\}$. This solves to $r_{13} = 0$, $r_{23} = 0$, $2x_1 \in \mathbb{Z}$, and $2x_2 \in \mathbb{Z}$.

Repeating this process for the other conditions gives the solution for the affine normalizer:

$N_A(pmm2) =$

$$\left\{ \begin{pmatrix} \pm 1 & 0 & 0 & x_1/2 \\ 0 & \pm 1 & 0 & x_2/2 \\ 0 & 0 & r_{33} & x_3 \end{pmatrix}, \begin{pmatrix} 0 & \pm 1 & 0 & x_1/2 \\ \pm 1 & 0 & 0 & x_2/2 \\ 0 & 0 & r_{33} & x_3 \end{pmatrix} : r_{33} \neq 0 \land r_{33}, x_3 \in \mathbb{R} \land x_1, x_2 \in \mathbb{Z} \right\}$$

The values of the $\pm 1$'s are independent.

Since $\{\{m_{100}|0\}\Lambda, \{m_{010}|0\}\Lambda\}$ is closed under conjugacy in $N_A(\Lambda)$ and generates $G/\Lambda$, it would have been sufficient to use the conditions relating to these.

### 4. Euclidean normalizers of subperiodic groups

Since the Euclidean normalizers are subgroups of the affine normalizers, in order to find the Euclidean normalizer of each subperiodic group $G$, we need only consider which elements of each affine normalizer preserve the distance metric $M$, i.e. which $\{a|\vec{t}\} \in N_A(G)$ satisfy $a^T M a = M$:

$$N_E(G) = \{\{a|\vec{t}\} \in N_A(G) : a^T M a = M\}$$

This expression comes from defining Euclidean length of a column vector $v$ as $d = \sqrt{v^T M v}$. For each symmetry group, there are constraints on the metric that must be satisfied in order for it to be consistent with the symmetry, e.g. in a cubic symmetry group $M$ is the identity matrix times the square of the lattice constant on the standard basis. When these constraints and no others are satisfied, the metric is considered "general". Special metrics that are more constrained than the general metric need to be considered for some of the layer groups.

### 4.1. Example: L1, *p*1

For *p*1, the most general type of metric is an oblique metric which is expressed on the standard crystal basis as:

$$\begin{pmatrix} a^2 & ab\cos\gamma & 0 \\ ab\cos\gamma & b^2 & 0 \\ 0 & 0 & 1 \end{pmatrix}$$

where $a$ and $b$ are the lengths of the unit cell vectors and $\gamma$ is the angle between them. The "1" in the final position indicates that a vector of unit length and orthogonal to the lattice vectors has been chosen as the arbitrary third basis vector. The affine normalizer of *p*1 is:

$$N_A(p1) = \left\{ \begin{pmatrix} n_{11} & n_{12} & r_{13} & x_1 \\ n_{21} & n_{22} & r_{23} & x_2 \\ 0 & 0 & r_{33} & x_3 \end{pmatrix} : \begin{vmatrix} n_{11} & n_{12} \\ n_{21} & n_{22} \end{vmatrix} = \pm 1 \wedge r_{33} \neq 0 \wedge r_{ij}, x_i \in \mathbb{R} \wedge n_{ij} \in \mathbb{Z} \right\}$$

We can find which of these satisfy $a^T M a = M$ by solving for the $n_{ij}$ and $r_{ij}$ values that solve

$$\begin{pmatrix} n_{11} & n_{21} & 0 \\ n_{12} & n_{22} & 0 \\ r_{13} & r_{23} & r_{33} \end{pmatrix} \begin{pmatrix} a^2 & ab\cos\gamma & 0 \\ ab\cos\gamma & b^2 & 0 \\ 0 & 0 & 1 \end{pmatrix} \begin{pmatrix} n_{11} & n_{12} & r_{13} \\ n_{21} & n_{22} & r_{23} \\ 0 & 0 & r_{33} \end{pmatrix} = \begin{pmatrix} a^2 & ab\cos\gamma & 0 \\ ab\cos\gamma & b^2 & 0 \\ 0 & 0 & 1 \end{pmatrix}$$

for all real values of $a$, $b$, and $\gamma$. The result is

$$N_E(p1) = \left\{ \begin{pmatrix} 1 & 0 & 0 & x_1 \\ 0 & 1 & 0 & x_2 \\ 0 & 0 & 1 & x_3 \end{pmatrix}, \begin{pmatrix} -1 & 0 & 0 & x_1 \\ 0 & -1 & 0 & x_2 \\ 0 & 0 & 1 & x_3 \end{pmatrix}, \begin{pmatrix} 1 & 0 & 0 & x_1 \\ 0 & 1 & 0 & x_2 \\ 0 & 0 & -1 & x_3 \end{pmatrix}, \begin{pmatrix} -1 & 0 & 0 & x_1 \\ 0 & -1 & 0 & x_2 \\ 0 & 0 & -1 & x_3 \end{pmatrix} : x_i \in \mathbb{R} \right\}$$

The Hermann-Mauguin symbol for this Euclidean normalizer group is $P^3 112/m$, which is similar to space group $P2/m$ but with continuous translational symmetry.

For the special case of *p*1 where the unit cell is square, $\gamma = 90°$ and $a = b$, the metric is

$$\begin{pmatrix} a^2 & 0 & 0 \\ 0 & a^2 & 0 \\ 0 & 0 & 1 \end{pmatrix}$$

For this special metric, the Euclidean normalizer group of *p*1 is $P^3 4/mmm$. The square lattice is invariant under four-fold rotation whereas the oblique lattice was not.

### 5. Normalizer Tables

Using the methods described in the present work, the affine and Euclidean normalizers of the subperiodic groups were derived and listed. To help validate our methods, the space group affine normalizers were computed and the results matched the listing found on the Bilbao Crystallographic Server. The normalizer tables for the Frieze, layer, and rod groups have been made to resemble the format of the space group and plane group normalizer tables given in the International Tables for Crystallography Vol. A. Each subperiodic group from the International Tables for Crystallography Vol. E (Kopský & Litvin, 2006) is listed sequentially in each table with a number and a Hermann-Mauguin symbol for each. Additionally, for the layer groups, special metric types were considered for cases where the results differ from a general metric.

After the columns specifying the subperiodic group, the Euclidean normalizer is given by a symbol and the basis vectors for the symbol in terms the subperiodic group's basis vectors. For cases where the Euclidean normalizer is a crystallographic group, this group is given with the standard symbol. For cases where it is not, a generalization of the standard symbol is given, $e.g.$ $N_E(\wp 6/mmm)$ is given as $\wp 12/mmm$. When the Euclidean normalizer contains continuous translation, a superscript number on the centering symbol denotes how many independent directions are continuous, $e.g.$ $N_E(\wp 4mm)$ contains continuous translation along the rod axis and thus is given as $\wp^1 8/mmm$. The standard choice of origin occurred for all Euclidean normalizers of the subperiodic groups except for three: layer groups 52, 62, and 64. In these cases, the symbol for $N_E(G)$ is given as $p4/mmm\ (mmm)$ to indicate that the origin is at a point with $mmm$ site symmetry rather than the standard origin.

Following the Euclidean normalizer, a set of additional generators are given which, when added to the generators of $G$, can be used to generate $N_E(G)$. The choice and description of additional generators was made to match the normalizer tables given in the International Tables for Crystallography Vol. A (Hahn, 2006; Koch et al. 2006).

The final column gives the index of $G$ in $N_E(G)$. This is factored in the same way as in the normalizer tables given in the International Tables for Crystallography Vol. A: the first number accounts for additional translational symmetry (we have factored this further into components that are perpendicular and parallel to the periodic subspace of $G$), the number in the second position is 2 if either inversion (rod and layer groups) or two-fold rotation (Frieze groups) is gained, otherwise it is 1, and the number in the final position is the remaining factor once the first two positions are accounted for. In the cases where continuous translation or continuous rotation is gained, ∞ appears in the factored index.

As is the case with the monoclinic and triclinic space groups, the affine normalizers of the subperiodic groups are not isomorphic to the Euclidean normalizers of any specialized metric. It is for this reason that Hermann-Mauguin symbols have not been given for the affine normalizers of the subperiodic groups. Instead, only the matrix representations of the elements of the affine normalizers of the subperiodic groups are listed.

The file "SubperiodicGroupNormalizersMachineReadableFiles.zip" contains the text files with matrix representatives of the elements of the affine and Euclidean normalizers for each subperiodic group. This file is given in the supplemental materials. The purpose of these text files is to allow the matrix representations of $N_E(G)$ and $N_A(G)$ for any subperiodic group $G$ to be easily parsed by a computer program.

Each normalizer listed in these text files starts with a single number on the first line giving the subperiodic group number. In the case of the Euclidean normalizers of the layer groups, the metric is also described on this line. The remaining lines give the augmented matrix representations of the elements of the normalizer flattened onto a single line. Thus for the Frieze group normalizers, there are nine entries on each of these lines which are to be partitioned sequentially into the rows of a 3x3 matrix. For the layer and rod group normalizers, there are sixteen entries on each of these lines which are to be partitioned sequentially into the rows of a 4x4 matrix. The matrix representations are given for the conventional setting only. Elements that differ by a translation of the conventional unit cell are listed by one representative, *i.e.* every matrix represents itself combined with any integer translation in the periodic subspace. There are non-numeric entries that are a letter followed by a number, such as "r1". These entries represent any real number if the letter is "r", any integer if the letter is "n", and any real number greater than or equal to zero and less than one if the letter is "x". The Mathematica file "Load Zipped SubperiodicGroupNormalizersMachineReadableFiles.nb" has also been provided in the supplemental materials. This file will parse the normalizers directly from the zip file.

**Acknowledgements**     We acknowledge support from the Penn State Center for Nanoscale Science through the NSF-MRSEC DMR #0820404. We also acknowledge DMR- 1210588.

|  |  | Total dimensionality |  |  |
|---|---|---|---|---|
|  |  | 1 | 2 | 3 |
| Periodic dimensions | 0 | 1D Point 2 | 2D Point 10 | 3D Point 32 |
|  | 1 | Line 2 | Frieze 7 | Rod 75 |
|  | 2 |  | Wallpaper 17 | Layer 80 |
|  | 3 |  |  | Space 230 |

**Figure 1:** Number of types of crystallographic symmetry groups across different values of $d$ and $p$.

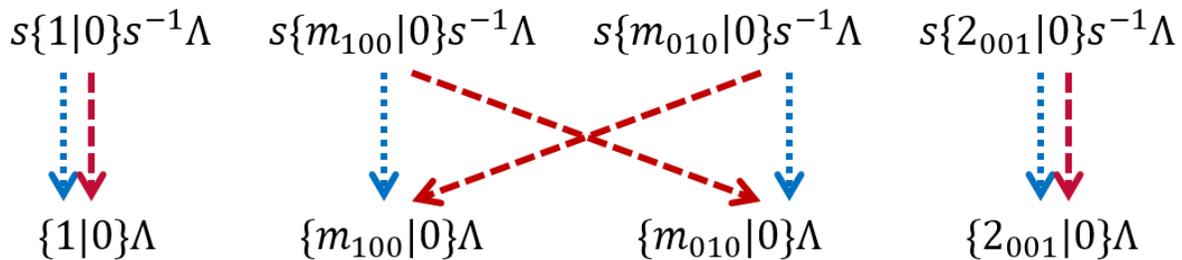

**Figure 2:** Possible permutations of $G/\Lambda$ by conjugation by an affine transformation when $G = pmm2$. The blue dotted arrows are the identity permutation, e.g. one of the many values of $s$ that would induce this permutation is $s = \{m_{001}|0\}$. The red dashed arrows are the permutation that swaps $\{m_{100}|0\}\Lambda$ and $\{m_{010}|0\}\Lambda$, e.g. $s = \{4_{001}|0\}$.

**Table 1** Euclidean normalizers of the Frieze groups

| | Frieze group $G$ | | Euclidean normalizer $N_E(G)$ | | Additional generators of $N_E(G)$ | | | |
|---|---|---|---|---|---|---|---|---|
| No. | Hermann-Mauguin symbol | Cell metric | Symbol | Basis vectors | Translations | Twofold rotation | Further generators | Index of $G$ in $N_E(G)$ |
| 1 | $p1$ | General | $p^22mm$ | $\varepsilon_1\mathbf{a}, \varepsilon_2\mathbf{b}$ | $x_1, 0; 0, x_2$ | $-x, -y$ | $-x, y$ | $(\infty_\parallel \cdot \infty_\perp) \cdot 2 \cdot 2$ |
| 2 | $p211$ | General | $p2mm$ | $\frac{1}{2}\mathbf{a}$ | $\frac{1}{2}, 0$ | | $-x, y$ | $2 \cdot 1 \cdot 2$ |
| 3 | $p1m1$ | | $p^12mm$ | $\frac{1}{2}\mathbf{a}, \varepsilon_2\mathbf{b}$ | $\frac{1}{2}, 0; 0, x_2$ | $-x, -y$ | | $(2 \cdot \infty_\perp) \cdot 2 \cdot 1$ |
| 4 | $p11m$ | | $p^12mm$ | $\varepsilon_1\mathbf{a}$ | $x_1, 0$ | $-x, -y$ | | $\infty_\parallel \cdot 2 \cdot 1$ |
| 5 | $p11g$ | | $p^12mm$ | $\varepsilon_1\mathbf{a}$ | $x_1, 0$ | $-x, -y$ | | $\infty_\parallel \cdot 2 \cdot 1$ |
| 6 | $p2mm$ | | $p2mm$ | $\frac{1}{2}\mathbf{a}$ | $\frac{1}{2}, 0$ | | | $2 \cdot 1 \cdot 1$ |
| 7 | $p2mg$ | | $p2mm$ | $\frac{1}{2}\mathbf{a}$ | $\frac{1}{2}, 0$ | | | $2 \cdot 1 \cdot 1$ |

**Table 2** Euclidean normalizers of the rod groups

| No. | Rod Group $G$ Hermann-Mauguin symbol | Euclidean normalizer $N_E(G)$ Symbol | Basis vectors | Additional generators of $N_E(G)$ Translations | Inversion | Further generators | Index of $G$ in $N_E(G)$ |
|---|---|---|---|---|---|---|---|
| 1 | $p1$ | $P^3\infty/mmm$ | $\varepsilon_1 a, \varepsilon_2 b, \varepsilon_3 c$ | $x_1,0,0; 0,x_2,0;$ $0,0,x_3$ | $(0,0,0)$ | Continuous rotation; $-x,y,z$ | $(\infty_\perp^2 \infty_\parallel) \cdot 2 \cdot \infty$ |
| 2 | $p\bar{1}$ | $p\infty/mmm$ | $a,b,\tfrac{1}{2}c$ | $0,0,\tfrac{1}{2}$ | | Continuous rotation; $-x,y,z$ | $2 \cdot 1 \cdot \infty$ |
| 3 | $p211$ | $p^1 mmm$ | $\varepsilon_1 a, b, \tfrac{1}{2}c$ | $x_1,0,0; 0,0,\tfrac{1}{2}$ | $(0,0,0)$ | $-x,-y,z$ | $(\infty_\perp 2) \cdot 2 \cdot 2$ |
| 4 | $pm11$ | $p^2 mmm$ | $a, \varepsilon_2 b, \varepsilon_3 c$ | $0,x_2,0;$ $0,0,x_3$ | $(0,0,0)$ | $-x,-y,z$ | $(\infty_\perp \infty_\parallel) \cdot 2 \cdot 2$ |
| 5 | $pc11$ | $p^2 mmm$ | $a, \varepsilon_2 b, \varepsilon_3 c$ | $0,x_2,0;$ $0,0,x_3$ | $(0,0,0)$ | $-x,-y,z$ | $(\infty_\perp \infty_\parallel) \cdot 2 \cdot 2$ |
| 6 | $p2/m11$ | $pmmm$ | $a,b,\tfrac{1}{2}c$ | $0,0,\tfrac{1}{2}$ | | $-x,-y,z$ | $2 \cdot 1 \cdot 2$ |
| 7 | $p2/c11$ | $pmmm$ | $a,b,\tfrac{1}{2}c$ | $0,0,\tfrac{1}{2}$ | | $-x,-y,z$ | $2 \cdot 1 \cdot 2$ |
| 8 | $p112$ | $p^1\infty/mmm$ | $a,b,\varepsilon_3 c$ | $0,0,x_3$ | $(0,0,0)$ | Continuous rotation; $-x,y,z$ | $(1\infty_\parallel) \cdot 2 \cdot \infty$ |
| 9 | $p112_1$ | $p^1\infty/mmm$ | $a,b,\varepsilon_3 c$ | $0,0,x_3$ | $(0,0,0)$ | Continuous rotation; $-x,y,z$ | $(1\infty_\parallel) \cdot 2 \cdot \infty$ |
| 10 | $p11m$ | $p^2\infty/mmm$ | $\varepsilon_1 a, \varepsilon_2 b, \tfrac{1}{2}c$ | $x_1,0,0; 0,x_2,0;$ $0,0,\tfrac{1}{2}$ | $(0,0,0)$ | Continuous rotation; $-x,y,z$ | $(\infty_\perp^2 2) \cdot 2 \cdot \infty$ |
| 11 | $p112/m$ | $p\infty/mmm$ | $a,b,\tfrac{1}{2}c$ | $0,0,\tfrac{1}{2}$ | | Continuous rotation; $-x,y,z$ | $2 \cdot 1 \cdot \infty$ |
| 12 | $p112_1/m$ | $p\infty/mmm$ | $a,b,\tfrac{1}{2}c$ | $0,0,\tfrac{1}{2}$ | | Continuous rotation; $-x,y,z$ | $2 \cdot 1 \cdot \infty$ |

| | | | | | | | | |
|---|---|---|---|---|---|---|---|---|
| 13 | $p222$ | $p4/mmm$ | $\boldsymbol{a},\boldsymbol{b},\frac{1}{2}\boldsymbol{c}$ | $0,0,\frac{1}{2}$ | $(0,0,0)$ | $-y,x,z$ | $2\cdot 2\cdot 2$ |
| 14 | $p222_1$ | $p4_2/mmc$ | $\boldsymbol{a},\boldsymbol{b},\frac{1}{2}\boldsymbol{c}$ | $0,0,\frac{1}{2}$ | $(0,0,0)$ | $-y,x,z$ | $2\cdot 2\cdot 2$ |
| 15 | $pmm2$ | $p^14/mmm$ | $\boldsymbol{a},\boldsymbol{b},\varepsilon_3\boldsymbol{c}$ | $0,0,x_3$ | $(0,0,0)$ | $-y,x,z$ | $(1\infty_\parallel)\cdot 2\cdot 2$ |
| 16 | $pcc2$ | $p^14/mmm$ | $\boldsymbol{a},\boldsymbol{b},\varepsilon_3\boldsymbol{c}$ | $0,0,x_3$ | $(0,0,0)$ | $-y,x,z$ | $(1\infty_\parallel)\cdot 2\cdot 2$ |
| 17 | $pmc2_1$ | $p^1mmm$ | $\boldsymbol{a},\boldsymbol{b},\varepsilon_3\boldsymbol{c}$ | $0,0,x_3$ | $(0,0,0)$ | | $(1\infty_\parallel)\cdot 2\cdot 1$ |
| 18 | $p2mm$ | $p^1mmm$ | $\varepsilon_1\boldsymbol{a},\boldsymbol{b},\frac{1}{2}\boldsymbol{c}$ | $x_1,0,0;0,0,\frac{1}{2}$ | $(0,0,0)$ | | $(\infty_\perp 2)\cdot 2\cdot 1$ |
| 19 | $p2cm$ | $p^1mmm$ | $\varepsilon_1\boldsymbol{a},\boldsymbol{b},\frac{1}{2}\boldsymbol{c}$ | $x_1,0,0;0,0,\frac{1}{2}$ | $(0,0,0)$ | | $(\infty_\perp 2)\cdot 2\cdot 1$ |
| 20 | $pmmm$ | $p4/mmm$ | $\boldsymbol{a},\boldsymbol{b},\frac{1}{2}\boldsymbol{c}$ | $0,0,\frac{1}{2}$ | | $-y,x,z$ | $2\cdot 1\cdot 2$ |
| 21 | $pccm$ | $p4/mmm$ | $\boldsymbol{a},\boldsymbol{b},\frac{1}{2}\boldsymbol{c}$ | $0,0,\frac{1}{2}$ | | $-y,x,z$ | $2\cdot 1\cdot 2$ |
| 22 | $pmcm$ | $pmmm$ | $\boldsymbol{a},\boldsymbol{b},\frac{1}{2}\boldsymbol{c}$ | $0,0,\frac{1}{2}$ | | | $2\cdot 1\cdot 1$ |
| 23 | $p4$ | $p^1\infty/mmm$ | $\boldsymbol{a},\boldsymbol{b},\varepsilon_3\boldsymbol{c}$ | $0,0,x_3$ | $(0,0,0)$ | Continuous rotation; $x,-y,-z$ | $(1\infty_\parallel)\cdot 2\cdot \infty$ |
| 24 | $p4_1$ | $p^1\infty 22$ | $\boldsymbol{a},\boldsymbol{b},\varepsilon_3\boldsymbol{c}$ | $0,0,x_3$ | | Continuous rotation; $x,-y,-z$ | $(1\infty_\parallel)\cdot 1\cdot \infty$ |
| 25 | $p4_2$ | $p^1\infty/mmm$ | $\boldsymbol{a},\boldsymbol{b},\varepsilon_3\boldsymbol{c}$ | $0,0,x_3$ | $(0,0,0)$ | Continuous rotation; $-x,y,z$ | $(1\infty_\parallel)\cdot 2\cdot \infty$ |
| 26 | $p4_3$ | $p^1\infty 22$ | $\boldsymbol{a},\boldsymbol{b},\varepsilon_3\boldsymbol{c}$ | $0,0,x_3$ | | Continuous rotation; $x,-y,-z$ | $(1\infty_\parallel)\cdot 1\cdot \infty$ |
| 27 | $p\bar{4}$ | $p\infty/mmm$ | $\boldsymbol{a},\boldsymbol{b},\frac{1}{2}\boldsymbol{c}$ | $0,0,\frac{1}{2}$ | $(0,0,0)$ | Continuous rotation; $-x,y,z$ | $2\cdot 2\cdot \infty$ |
| 28 | $p4/m$ | $p\infty/mmm$ | $\boldsymbol{a},\boldsymbol{b},\frac{1}{2}\boldsymbol{c}$ | $0,0,\frac{1}{2}$ | | Continuous rotation; $-x,y,z$ | $2\cdot 1\cdot \infty$ |
| 29 | $p4_2/m$ | $p\infty/mmm$ | $\boldsymbol{a},\boldsymbol{b},\frac{1}{2}\boldsymbol{c}$ | $0,0,\frac{1}{2}$ | | Continuous rotation; $-x,y,z$ | $2\cdot 1\cdot \infty$ |
| 30 | $p422$ | $p8/mmm$ | $\boldsymbol{a},\boldsymbol{b},\frac{1}{2}\boldsymbol{c}$ | $0,0,\frac{1}{2}$ | $(0,0,0)$ | $\frac{1}{\sqrt{2}}x-\frac{1}{\sqrt{2}}y,\frac{1}{\sqrt{2}}x+\frac{1}{\sqrt{2}}y,z$ | $2\cdot 2\cdot 2$ |
| 31 | $p4_122$ | $p8_222$ | $\boldsymbol{a},\boldsymbol{b},\frac{1}{2}\boldsymbol{c}$ | $0,0,\frac{1}{2}$ | | $\frac{1}{\sqrt{2}}x-\frac{1}{\sqrt{2}}y,\frac{1}{\sqrt{2}}x+\frac{1}{\sqrt{2}}y,z$ | $2\cdot 1\cdot 2$ |

| # | | | | | | | |
|---|---|---|---|---|---|---|---|
| 32 | $p4_222$ | $p8_4/mmc$ | $\boldsymbol{a},\boldsymbol{b},\frac{1}{2}\boldsymbol{c}$ | $0,0,\frac{1}{2}$ | $(0,0,0)$ | $\frac{1}{\sqrt{2}}x-\frac{1}{\sqrt{2}}y,\frac{1}{\sqrt{2}}x+\frac{1}{\sqrt{2}}y,z$ | $2\cdot2\cdot2$ |
| 33 | $p4_322$ | $p8_622$ | $\boldsymbol{a},\boldsymbol{b},\frac{1}{2}\boldsymbol{c}$ | $0,0,\frac{1}{2}$ | | $\frac{1}{\sqrt{2}}x-\frac{1}{\sqrt{2}}y,\frac{1}{\sqrt{2}}x+\frac{1}{\sqrt{2}}y,z$ | $2\cdot1\cdot2$ |
| 34 | $p4mm$ | $p^18/mmm$ | $\boldsymbol{a},\boldsymbol{b},\varepsilon_3\boldsymbol{c}$ | $0,0,x_3$ | $(0,0,0)$ | $\frac{1}{\sqrt{2}}x-\frac{1}{\sqrt{2}}y,\frac{1}{\sqrt{2}}x+\frac{1}{\sqrt{2}}y,z$ | $(1\infty_\parallel)\cdot2\cdot2$ |
| 35 | $p4_2mc$ | $p^14/mmm$ | $\boldsymbol{a},\boldsymbol{b},\varepsilon_3\boldsymbol{c}$ | $0,0,x_3$ | $(0,0,0)$ | | $(\infty_\parallel 1)\cdot2\cdot1$ |
| 36 | $p4cc$ | $p^18/mmm$ | $\boldsymbol{a},\boldsymbol{b},\varepsilon_3\boldsymbol{c}$ | $0,0,x_3$ | $(0,0,0)$ | $\frac{1}{\sqrt{2}}x-\frac{1}{\sqrt{2}}y,\frac{1}{\sqrt{2}}x+\frac{1}{\sqrt{2}}y,z$ | $(1\infty_\parallel)\cdot2\cdot2$ |
| 37 | $p\bar{4}2m$ | $p4/mmm$ | $\boldsymbol{a},\boldsymbol{b},\frac{1}{2}\boldsymbol{c}$ | $0,0,\frac{1}{2}$ | $(0,0,0)$ | | $2\cdot2\cdot1$ |
| 38 | $p\bar{4}2c$ | $p4/mmm$ | $\boldsymbol{a},\boldsymbol{b},\frac{1}{2}\boldsymbol{c}$ | $0,0,\frac{1}{2}$ | $(0,0,0)$ | | $2\cdot2\cdot1$ |
| 39 | $p4/mmm$ | $p8/mmm$ | $\boldsymbol{a},\boldsymbol{b},\frac{1}{2}\boldsymbol{c}$ | $0,0,\frac{1}{2}$ | | $\frac{1}{\sqrt{2}}x-\frac{1}{\sqrt{2}}y,\frac{1}{\sqrt{2}}x+\frac{1}{\sqrt{2}}y,z$ | $2\cdot1\cdot2$ |
| 40 | $p4/mcc$ | $p8/mmm$ | $\boldsymbol{a},\boldsymbol{b},\frac{1}{2}\boldsymbol{c}$ | $0,0,\frac{1}{2}$ | | $\frac{1}{\sqrt{2}}x-\frac{1}{\sqrt{2}}y,\frac{1}{\sqrt{2}}x+\frac{1}{\sqrt{2}}y,z$ | $2\cdot1\cdot2$ |
| 41 | $p4_2/mmc$ | $p4/mmm$ | $\boldsymbol{a},\boldsymbol{b},\frac{1}{2}\boldsymbol{c}$ | $0,0,\frac{1}{2}$ | | | $2\cdot1\cdot1$ |
| 42 | $p3$ | $p^1\infty/mmm$ | $\boldsymbol{a},\boldsymbol{b},\varepsilon_3\boldsymbol{c}$ | $0,0,x_3$ | $(0,0,0)$ | Continuous rotation; $-x,y,z$ | $(1\infty_\parallel)\cdot2\cdot\infty$ |
| 43 | $p3_1$ | $p^1\infty22$ | $\boldsymbol{a},\boldsymbol{b},\varepsilon_3\boldsymbol{c}$ | $0,0,x_3$ | | Continuous rotation; $x,-y,-z$ | $(1\infty_\parallel)\cdot1\cdot\infty$ |
| 44 | $p3_2$ | $p^1\infty22$ | $\boldsymbol{a},\boldsymbol{b},\varepsilon_3\boldsymbol{c}$ | $0,0,x_3$ | | Continuous rotation; $x,-y,-z$ | $(1\infty_\parallel)\cdot1\cdot\infty$ |
| 45 | $p\bar{3}$ | $p\infty/mmm$ | $\boldsymbol{a},\boldsymbol{b},\frac{1}{2}\boldsymbol{c}$ | $0,0,\frac{1}{2}$ | | Continuous rotation; $-x,y,z$ | $2\cdot1\cdot\infty$ |
| 46 | $p312$ | $p6/mmm$ | $\frac{1}{3}(2\boldsymbol{a}+\boldsymbol{b}),\frac{1}{3}(-\boldsymbol{a}+\boldsymbol{b}),\frac{1}{2}\boldsymbol{c}$ | $0,0,\frac{1}{2}$ | $(0,0,0)$ | $-x,-y,z$ | $2\cdot2\cdot2$ |
| 47 | $p3_112$ | $p6_222$ | $\frac{1}{3}(2\boldsymbol{a}+\boldsymbol{b}),\frac{1}{3}(-\boldsymbol{a}+\boldsymbol{b}),\frac{1}{2}\boldsymbol{c}$ | $0,0,\frac{1}{2}$ | | $-x,-y,z$ | $2\cdot1\cdot2$ |
| 48 | $p3_212$ | $p6_422$ | $\frac{1}{3}(2\boldsymbol{a}+\boldsymbol{b}),\frac{1}{3}(-\boldsymbol{a}+\boldsymbol{b}),\frac{1}{2}\boldsymbol{c}$ | $0,0,\frac{1}{2}$ | | $-x,-y,z$ | $2\cdot1\cdot2$ |
| 49 | $p3m1$ | $p^16/mmm$ | $\boldsymbol{a},\boldsymbol{b},\varepsilon_3\boldsymbol{c}$ | $0,0,x_3$ | $(0,0,0)$ | $-x,-y,z$ | $(1\infty_\parallel)\cdot2\cdot2$ |
| 50 | $p3c1$ | $p^16/mmm$ | $\boldsymbol{a},\boldsymbol{b},\varepsilon_3\boldsymbol{c}$ | $0,0,x_3$ | $(0,0,0)$ | $-x,-y,z$ | $(1\infty_\parallel)\cdot2\cdot2$ |
| 51 | $p\bar{3}m1$ | $p6/mmm$ | $\boldsymbol{a},\boldsymbol{b},\frac{1}{2}\boldsymbol{c}$ | $0,0,\frac{1}{2}$ | | $-x,-y,z$ | $2\cdot1\cdot2$ |

| | | | | | | | | |
|---|---|---|---|---|---|---|---|---|
| 52 | $p\bar{3}c1$ | $p6/mmm$ | $\boldsymbol{a},\boldsymbol{b},\frac{1}{2}\boldsymbol{c}$ | $0,0,\frac{1}{2}$ | | $-x,-y,z$ | $2\cdot 1\cdot 2$ |
| 53 | $p6$ | $p^1\infty/mmm$ | $\boldsymbol{a},\boldsymbol{b},\varepsilon_3\boldsymbol{c}$ | $0,0,x_3$ | $(0,0,0)$ | Continuous rotation; $-x,y,z$ | $(1\infty_\parallel)\cdot 2\cdot \infty$ |
| 54 | $p6_1$ | $p^1\infty 22$ | $\boldsymbol{a},\boldsymbol{b},\varepsilon_3\boldsymbol{c}$ | $0,0,x_3$ | | Continuous rotation; $x,-y,-z$ | $(1\infty_\parallel)\cdot 1\cdot \infty$ |
| 55 | $p6_2$ | $p^1\infty 22$ | $\boldsymbol{a},\boldsymbol{b},\varepsilon_3\boldsymbol{c}$ | $0,0,x_3$ | | Continuous rotation; $x,-y,-z$ | $(1\infty_\parallel)\cdot 1\cdot \infty$ |
| 56 | $p6_3$ | $p^1\infty/mmm$ | $\boldsymbol{a},\boldsymbol{b},\varepsilon_3\boldsymbol{c}$ | $0,0,x_3$ | $(0,0,0)$ | Continuous rotation; $-x,y,z$ | $(1\infty_\parallel)\cdot 2\cdot \infty$ |
| 57 | $p6_4$ | $p^1\infty 22$ | $\boldsymbol{a},\boldsymbol{b},\varepsilon_3\boldsymbol{c}$ | $0,0,x_3$ | | Continuous rotation; $x,-y,-z$ | $(1\infty_\parallel)\cdot 1\cdot \infty$ |
| 58 | $p6_5$ | $p^1\infty 22$ | $\boldsymbol{a},\boldsymbol{b},\varepsilon_3\boldsymbol{c}$ | $0,0,x_3$ | | Continuous rotation; $x,-y,-z$ | $(1\infty_\parallel)\cdot 1\cdot \infty$ |
| 59 | $p\bar{6}$ | $p\infty/mmm$ | $\boldsymbol{a},\boldsymbol{b},\frac{1}{2}\boldsymbol{c}$ | $0,0,\frac{1}{2}$ | $(0,0,0)$ | Continuous rotation; $-x,y,z$ | $2\cdot 2\cdot \infty$ |
| 60 | $p6/m$ | $p\infty/mmm$ | $\boldsymbol{a},\boldsymbol{b},\frac{1}{2}\boldsymbol{c}$ | $0,0,\frac{1}{2}$ | | Continuous rotation; $-x,y,z$ | $2\cdot 1\cdot \infty$ |
| 61 | $p6_3/m$ | $p\infty/mmm$ | $\boldsymbol{a},\boldsymbol{b},\frac{1}{2}\boldsymbol{c}$ | $0,0,\frac{1}{2}$ | | Continuous rotation; $-x,y,z$ | $2\cdot 1\cdot \infty$ |
| 62 | $p622$ | $p12/mmm$ | $\boldsymbol{a},\boldsymbol{b},\frac{1}{2}\boldsymbol{c}$ | $0,0,\frac{1}{2}$ | $(0,0,0)$ | $\frac{1}{\sqrt{3}}x-\frac{2}{\sqrt{3}}y,\frac{2}{\sqrt{3}}x-\frac{1}{\sqrt{3}}y,z$ | $2\cdot 2\cdot 2$ |
| 63 | $p6_122$ | $p12_222$ | $\boldsymbol{a},\boldsymbol{b},\frac{1}{2}\boldsymbol{c}$ | $0,0,\frac{1}{2}$ | | $\frac{1}{\sqrt{3}}x-\frac{2}{\sqrt{3}}y,\frac{2}{\sqrt{3}}x-\frac{1}{\sqrt{3}}y,z$ | $2\cdot 1\cdot 2$ |
| 64 | $p6_222$ | $p12_422$ | $\boldsymbol{a},\boldsymbol{b},\frac{1}{2}\boldsymbol{c}$ | $0,0,\frac{1}{2}$ | | $\frac{1}{\sqrt{3}}x-\frac{2}{\sqrt{3}}y,\frac{2}{\sqrt{3}}x-\frac{1}{\sqrt{3}}y,z$ | $2\cdot 1\cdot 2$ |
| 65 | $p6_322$ | $p12_6/mmc$ | $\boldsymbol{a},\boldsymbol{b},\frac{1}{2}\boldsymbol{c}$ | $0,0,\frac{1}{2}$ | $(0,0,0)$ | $\frac{1}{\sqrt{3}}x-\frac{2}{\sqrt{3}}y,\frac{2}{\sqrt{3}}x-\frac{1}{\sqrt{3}}y,z$ | $2\cdot 2\cdot 2$ |
| 66 | $p6_422$ | $p12_822$ | $\boldsymbol{a},\boldsymbol{b},\frac{1}{2}\boldsymbol{c}$ | $0,0,\frac{1}{2}$ | | $\frac{1}{\sqrt{3}}x-\frac{2}{\sqrt{3}}y,\frac{2}{\sqrt{3}}x-\frac{1}{\sqrt{3}}y,z$ | $2\cdot 1\cdot 2$ |
| 67 | $p6_522$ | $p12_{10}22$ | $\boldsymbol{a},\boldsymbol{b},\frac{1}{2}\boldsymbol{c}$ | $0,0,\frac{1}{2}$ | | $\frac{1}{\sqrt{3}}x-\frac{2}{\sqrt{3}}y,\frac{2}{\sqrt{3}}x-\frac{1}{\sqrt{3}}y,z$ | $2\cdot 1\cdot 2$ |

| | | | | | | | | |
|---|---|---|---|---|---|---|---|---|
| 68 | $p6mm$ | $p^112/mmm$ | $\boldsymbol{a}, \boldsymbol{b}, \varepsilon_3 \boldsymbol{c}$ | $0,0,x_3$ | $(0,0,0)$ | $\frac{1}{\sqrt{3}}x - \frac{2}{\sqrt{3}}y, \frac{2}{\sqrt{3}}x - \frac{1}{\sqrt{3}}y, z$ | $(1\infty_{\parallel}) \cdot 2 \cdot 2$ |
| 69 | $p6cc$ | $p^112/mmm$ | $\boldsymbol{a}, \boldsymbol{b}, \varepsilon_3 \boldsymbol{c}$ | $0,0,x_3$ | $(0,0,0)$ | $\frac{1}{\sqrt{3}}x - \frac{2}{\sqrt{3}}y, \frac{2}{\sqrt{3}}x - \frac{1}{\sqrt{3}}y, z$ | $(1\infty_{\parallel}) \cdot 2 \cdot 2$ |
| 70 | $p6_3mc$ | $p^16/mmm$ | $\boldsymbol{a}, \boldsymbol{b}, \varepsilon_3 \boldsymbol{c}$ | $0,0,x_3$ | $(0,0,0)$ | | $(1\infty_{\parallel}) \cdot 2 \cdot 1$ |
| 71 | $p\bar{6}m2$ | $p6/mmm$ | $\boldsymbol{a}, \boldsymbol{b}, \frac{1}{2}\boldsymbol{c}$ | $0,0,\frac{1}{2}$ | $(0,0,0)$ | | $2 \cdot 2 \cdot 1$ |
| 72 | $p\bar{6}c2$ | $p6/mmm$ | $\boldsymbol{a}, \boldsymbol{b}, \frac{1}{2}\boldsymbol{c}$ | $0,0,\frac{1}{2}$ | $(0,0,0)$ | | $2 \cdot 2 \cdot 1$ |
| 73 | $p6/mmm$ | $p12/mmm$ | $\boldsymbol{a}, \boldsymbol{b}, \frac{1}{2}\boldsymbol{c}$ | $0,0,\frac{1}{2}$ | | $\frac{1}{\sqrt{3}}x - \frac{2}{\sqrt{3}}y, \frac{2}{\sqrt{3}}x - \frac{1}{\sqrt{3}}y, z$ | $2 \cdot 1 \cdot 2$ |
| 74 | $p6/mcc$ | $p12/mmm$ | $\boldsymbol{a}, \boldsymbol{b}, \frac{1}{2}\boldsymbol{c}$ | $0,0,\frac{1}{2}$ | | $\frac{1}{\sqrt{3}}x - \frac{2}{\sqrt{3}}y, \frac{2}{\sqrt{3}}x - \frac{1}{\sqrt{3}}y, z$ | $2 \cdot 1 \cdot 2$ |
| 75 | $p6/mmc$ | $p6/mmm$ | $\boldsymbol{a}, \boldsymbol{b}, \frac{1}{2}\boldsymbol{c}$ | $0,0,\frac{1}{2}$ | | | $2 \cdot 1 \cdot 1$ |

**Table 3** Euclidean normalizers of the layer groups

| Layer Group $G$ | | | Euclidean normalizer $N_E(G)$ | | Additional generators of $N_E(G)$ | | | Index of $G$ in $N_E(G)$ |
|---|---|---|---|---|---|---|---|---|
| | Hermann-Mauguin symbol | Cell Metric | Symbol | Basis vectors | Translations | Inversion | Further generators | |
| 1 | $p1$ | General Metric | $P^3 112/m$ | $\varepsilon_1 \boldsymbol{a}, \varepsilon_2 \boldsymbol{b}, \varepsilon_3 \boldsymbol{c}$ | $x_1, 0, 0; 0, x_2, 0; 0, 0, x_3$ | $(0,0,0)$ | $-x, -y, z$ | $(\infty_\perp \infty_\parallel^2) \cdot 2 \cdot 2$ |
| | | $a < b, \gamma = 90°$ | $P^3 mmm$ | $\varepsilon_1 \boldsymbol{a}, \varepsilon_2 \boldsymbol{b}, \varepsilon_3 \boldsymbol{c}$ | $x_1, 0, 0; 0, x_2, 0; 0, 0, x_3$ | $(0,0,0)$ | $-x, -y, z; -x, y, -z$ | $(\infty_\perp \infty_\parallel^2) \cdot 2 \cdot 4$ |
| | | $a = b, \gamma = 90°$ | $P^3 4/mmm$ | $\varepsilon_1 \boldsymbol{a}, \varepsilon_2 \boldsymbol{b}, \varepsilon_3 \boldsymbol{c}$ | $x_1, 0, 0; 0, x_2, 0; 0, 0, x_3$ | $(0,0,0)$ | $-y, x, z; -x, -y, z; -x, y, -z$ | $(\infty_\perp \infty_\parallel^2) \cdot 2 \cdot 8$ |
| | | $2 \cos \gamma = \frac{-a}{b}, 90° < \gamma < 120°$ | $C^3 mmm$ | $\varepsilon_1 \boldsymbol{a}, \varepsilon_2 \left(\frac{1}{2}\boldsymbol{a} + \boldsymbol{b}\right), \varepsilon_3 \boldsymbol{c}$ | $x_1, 0, 0; 0, x_2, 0; 0, 0, x_3$ | $(0,0,0)$ | $-x, -y, z; -x + y, y, z$ | $(\infty_\perp \infty_\parallel^2) \cdot 2 \cdot 4$ |
| | | $a = b, 90° < \gamma < 120°$ | $C^3 mmm$ | $\frac{\varepsilon_1}{2}(\boldsymbol{a}+\boldsymbol{b}), \frac{\varepsilon_2}{2}(\boldsymbol{b}-\boldsymbol{a}), \varepsilon_3 \boldsymbol{c}$ | $x_1, 0, 0; 0, x_2, 0; 0, 0, x_3$ | $(0,0,0)$ | $-x, -y, z; -y, -x, z$ | $(\infty_\perp \infty_\parallel^2) \cdot 2 \cdot 4$ |
| | | $a = b, \gamma = 120°$ | $P^3 6/mmm$ | $\varepsilon_1 \boldsymbol{a}, \varepsilon_2 \boldsymbol{b}, \varepsilon_3 \boldsymbol{c}$ | $x_1, 0, 0; 0, x_2, 0; 0, 0, x_3$ | $(0,0,0)$ | $-y, x$; $-y, z; -x, -y, z; y, x, z$ | $(\infty_\perp \infty_\parallel^2) \cdot 2 \cdot 12$ |
| 2 | $p\bar{1}$ | General Metric | $P^1 112/m$ | $\frac{1}{2}\boldsymbol{a}, \frac{1}{2}\boldsymbol{b}, \varepsilon_3 \boldsymbol{c}$ | $\frac{1}{2}, 0, 0; 0, \frac{1}{2}, 0; 0, 0, x_3$ | | $-x, -y, z$ | $(\infty_\perp 4) \cdot 1 \cdot 2$ |
| | | $a < b, \gamma = 90°$ | $P^1 mmm$ | $\frac{1}{2}\boldsymbol{a}, \frac{1}{2}\boldsymbol{b}, \varepsilon_3 \boldsymbol{c}$ | $\frac{1}{2}, 0, 0; 0, \frac{1}{2}, 0; 0, 0, x_3$ | | $-x, -y, z; -x, y, -z$ | $(\infty_\perp 4) \cdot 1 \cdot 4$ |
| | | $a = b, \gamma = 90°$ | $P^1 4/mmm$ | $\frac{1}{2}\boldsymbol{a}, \frac{1}{2}\boldsymbol{b}, \varepsilon_3 \boldsymbol{c}$ | $\frac{1}{2}, 0, 0; 0, \frac{1}{2}, 0; 0, 0, x_3$ | | $-y, x, z; -x, -y, z; -x, y, -z$ | $(\infty_\perp 4) \cdot 1 \cdot 8$ |
| | | $2 \cos \gamma = \frac{-a}{b}, 90° < \gamma < 120°$ | $C^1 mmm$ | $\frac{1}{2}\boldsymbol{a}, \frac{1}{2}\boldsymbol{a} + \boldsymbol{b}, \varepsilon_3 \boldsymbol{c}$ | $\frac{1}{2}, 0, 0; 0, \frac{1}{2}, 0; 0, 0, x_3$ | | $-x, -y, z; -x + y, y, z$ | $(\infty_\perp 4) \cdot 1 \cdot 4$ |
| | | $a = b, 90° < \gamma < 120°$ | $C^1 mmm$ | $\frac{1}{2}(\boldsymbol{a}+\boldsymbol{b}), \frac{1}{2}(\boldsymbol{b}-\boldsymbol{a}), \varepsilon_3 \boldsymbol{c}$ | $\frac{1}{2}, 0, 0; 0, \frac{1}{2}, 0; 0, 0, x_3$ | | $-x, -y, z; -y, -x, z$ | $(\infty_\perp 4) \cdot 1 \cdot 4$ |
| | | $a = b, \gamma = 120°$ | $P^1 6/mmm$ | $\varepsilon_3 \boldsymbol{c}$ | $0, 0, x_3$ | | $-y, x$; $-y, z; -x, -y, z; y, x, z$ | $(\infty_\perp 1) \cdot 1 \cdot 12$ |
| 3 | $p112$ | General Metric | $P^1 112/m$ | $\frac{1}{2}\boldsymbol{a}, \frac{1}{2}\boldsymbol{b}, \varepsilon_3 \boldsymbol{c}$ | $\frac{1}{2}, 0, 0; 0, \frac{1}{2}, 0; 0, 0, x_3$ | $(0,0,0)$ | | $(\infty_\perp 4) \cdot 2 \cdot 1$ |
| | | $a < b, \gamma = 90°$ | $P^1 mmm$ | $\frac{1}{2}\boldsymbol{a}, \frac{1}{2}\boldsymbol{b}, \varepsilon_3 \boldsymbol{c}$ | $\frac{1}{2}, 0, 0; 0, \frac{1}{2}, 0; 0, 0, x_3$ | $(0,0,0)$ | $-x, y, -z$ | $(\infty_\perp 4) \cdot 2 \cdot 2$ |
| | | $a = b, \gamma = 90°$ | $P^1 4/mmm$ | $\frac{1}{2}\boldsymbol{a}, \frac{1}{2}\boldsymbol{b}, \varepsilon_3 \boldsymbol{c}$ | $\frac{1}{2}, 0, 0; 0, \frac{1}{2}, 0; 0, 0, x_3$ | $(0,0,0)$ | $-y, x, z; -x, y, -z$ | $(\infty_\perp 4) \cdot 2 \cdot 4$ |
| | | $2 \cos \gamma = \frac{-a}{b}, 90° < \gamma < 120°$ | $C^1 mmm$ | $\frac{1}{2}\boldsymbol{a}, \frac{1}{2}\boldsymbol{a} + \boldsymbol{b}, \varepsilon_3 \boldsymbol{c}$ | $\frac{1}{2}, 0, 0; 0, \frac{1}{2}, 0; 0, 0, x_3$ | $(0,0,0)$ | $-x + y, y, z$ | $(\infty_\perp 4) \cdot 2 \cdot 2$ |
| | | $a = b, 90° < \gamma < 120°$ | $C^1 mmm$ | $\frac{1}{2}(\boldsymbol{a}+\boldsymbol{b}), \frac{1}{2}(\boldsymbol{b}-\boldsymbol{a}), \varepsilon_3 \boldsymbol{c}$ | $\frac{1}{2}, 0, 0; 0, \frac{1}{2}, 0; 0, 0, x_3$ | $(0,0,0)$ | $-y, -x, -z$ | $(\infty_\perp 4) \cdot 2 \cdot 2$ |
| | | $a = b, \gamma = 120°$ | $P^1 6/mmm$ | $\varepsilon_3 \boldsymbol{c}$ | $0, 0, x_3$ | $(0,0,0)$ | $-y, x - y, z; y, x, -z$ | $(\infty_\perp 1) \cdot 2 \cdot 6$ |

| | | | | | | | | | |
|---|---|---|---|---|---|---|---|---|---|
| 4 | p11m | General Metric | $P^3112/m$ | $\varepsilon_1 a, \varepsilon_2 b, \varepsilon_3 c$ | $x_1,0,0; 0,x_2,0; 0,0,x_3$ | $(0,0,0)$ | | | $(\infty_\perp \infty_\parallel^2) \cdot 2 \cdot 1$ |
| | | $a < b, \gamma = 90°$ | $P^3mmm$ | $\varepsilon_1 a, \varepsilon_2 b, \varepsilon_3 c$ | $x_1,0,0; 0,x_2,0; 0,0,x_3$ | $(0,0,0)$ | | $-x,y,-z$ | $(\infty_\perp \infty_\parallel^2) \cdot 2 \cdot 2$ |
| | | $a = b, \gamma = 90°$ | $P^3 4/mmm$ | $\varepsilon_1 a, \varepsilon_2 b, \varepsilon_3 c$ | $x_1,0,0; 0,x_2,0; 0,0,x_3$ | $(0,0,0)$ | | $-y,x,z; -x,y,-z$ | $(\infty_\perp \infty_\parallel^2) \cdot 2 \cdot 4$ |
| | | $2\cos\gamma = \frac{-a}{b}, 90°<\gamma<120°$ | $C^3mmm$ | $\varepsilon_1 a, \varepsilon_2(\frac{1}{2}a + b), \varepsilon_3 c$ | $x_1,0,0; 0,x_2,0; 0,0,x_3$ | $(0,0,0)$ | | $-x+y,y,z$ | $(\infty_\perp \infty_\parallel^2) \cdot 2 \cdot 2$ |
| | | $a = b, 90°<\gamma<120°$ | $C^3mmm$ | $\frac{\varepsilon_1}{2}(a+b), \frac{\varepsilon_2}{2}(b-a), \varepsilon_3 c$ | $x_1,0,0; 0,x_2,0; 0,0,x_3$ | $(0,0,0)$ | | $-y,-x,-z$ | $(\infty_\perp \infty_\parallel^2) \cdot 2 \cdot 2$ |
| | | $a = b, \gamma = 120°$ | $P^3 6/mmm$ | $\varepsilon_1 a, \varepsilon_2 b, \varepsilon_3 c$ | $x_1,0,0; 0,x_2,0; 0,0,x_3$ | $(0,0,0)$ | | $-y,x-y,z; y,x,-z$ | $(\infty_\perp \infty_\parallel^2) \cdot 2 \cdot 6$ |
| 5 | p11a | General Metric | $p^2 112/m$ | $\varepsilon_1 a, \varepsilon_2 b, c$ | $x_1,0,0; 0,x_2,0$ | $(0,0,0)$ | | | $(1\infty_\parallel^2) \cdot 2 \cdot 1$ |
| | | $a < b, \gamma = 90°$ | $p^2 mmm$ | $\varepsilon_1 a, \varepsilon_2 b, c$ | $x_1,0,0; 0,x_2,0$ | $(0,0,0)$ | | $-x,y,-z$ | $(1\infty_\parallel^2) \cdot 2 \cdot 2$ |
| | | $a = b, \gamma = 90°$ | $c^2 mmm$ | $\varepsilon_1 a, \varepsilon_2 b, c$ | $x_1,0,0; 0,x_2,0$ | $(0,0,0)$ | | $-x,y,-z$ | $(1\infty_\parallel^2) \cdot 2 \cdot 2$ |
| | | $2\cos\gamma = \frac{-a}{b}, 90°<\gamma<120°$ | $c^2 mmm$ | $\varepsilon_1 a, \varepsilon_2(\frac{1}{2}a + b), c$ | | | | $x-y,-y,z$ | $(1\infty_\parallel^2) \cdot 2 \cdot 2$ |
| | | $a = b, 90°<\gamma<120°$ | $c^2 mmm$ | $\frac{\varepsilon_1}{2}(a+b), \frac{\varepsilon_2}{2}(b-a), c$ | $x_1,0,0; 0,x_2,0$ | $(0,0,0)$ | | $-y,-x,z$ | $(1\infty_\parallel^2) \cdot 2 \cdot 2$ |
| | | $a = b, \gamma = 120°$ | $c^2 mmm$ | $\varepsilon_1 a, \varepsilon_2(\frac{1}{2}a + b), c$ | $x_1,0,0; 0,x_2,0$ | $(0,0,0)$ | | $x-y,-y,-z$ | $(1\infty_\parallel^2) \cdot 2 \cdot 2$ |
| | | | | | $x_1,0,0; 0,x_2,0$ | $(0,0,0)$ | | | |
| 6 | p112/m | General Metric | $p112/m$ | $\frac{1}{2}a, \frac{1}{2}b, c$ | $\frac{1}{2},0,0; 0,\frac{1}{2},0$ | | | | $4 \cdot 1 \cdot 1$ |
| | | $a < b, \gamma = 90°$ | $pmmm$ | $\frac{1}{2}a, \frac{1}{2}b, c$ | $\frac{1}{2},0,0; 0,\frac{1}{2},0$ | | | $-x,y,-z$ | $4 \cdot 1 \cdot 2$ |
| | | $a = b, \gamma = 90°$ | $p4/mmm$ | $\frac{1}{2}a, \frac{1}{2}b, c$ | $\frac{1}{2},0,0; 0,\frac{1}{2},0$ | | | $-y,x,z; -x,y,-z$ | $4 \cdot 1 \cdot 4$ |
| | | $2\cos\gamma = \frac{-a}{b}, 90°<\gamma<120°$ | $cmmm$ | $\frac{1}{2}a, \frac{1}{2}a + b, c$ | $\frac{1}{2},0,0; 0,\frac{1}{2},0$ | | | $-x+y,y,z$ | $4 \cdot 1 \cdot 2$ |
| | | $a = b, 90°<\gamma<120°$ | $cmmm$ | $\frac{1}{2}(a+b), \frac{1}{2}(b-a), c$ | $\frac{1}{2},0,0; 0,\frac{1}{2},0$ | | | $-y,-x,-z$ | $4 \cdot 1 \cdot 2$ |
| | | $a = b, \gamma = 120°$ | $p6/mmm$ | $a, b, c$ | | | | $-y,x,z; -x+y,y,z$ | $1 \cdot 1 \cdot 3$ |
| 7 | p112/a | General Metric | $p112/m$ | $\frac{1}{2}a, \frac{1}{2}b, c$ | $\frac{1}{2},0,0; 0,\frac{1}{2},0$ | | | | $4 \cdot 1 \cdot 1$ |
| | | $a < b, \gamma = 90°$ | $pmmm$ | $\frac{1}{2}a, \frac{1}{2}b, c$ | $\frac{1}{2},0,0; 0,\frac{1}{2},0$ | | | $-x,y,-z$ | $4 \cdot 1 \cdot 2$ |
| | | $a = b, \gamma = 90°$ | $cmmm$ | $\frac{1}{2}a, \frac{1}{2}b, c$ | $\frac{1}{2},0,0; 0,\frac{1}{2},0$ | | | $-x,y,-z$ | $4 \cdot 1 \cdot 2$ |
| | | $2\cos\gamma = \frac{-a}{b}, 90°<\gamma<120°$ | $cmmm$ | $\frac{1}{2}a, \frac{1}{2}a + b, c$ | $\frac{1}{2},0,0; 0,\frac{1}{2},0$ | | | $x-y,-y,z$ | $4 \cdot 1 \cdot 2$ |
| | | $a = b, 90°<\gamma<120°$ | $cmmm$ | $\frac{1}{2}(a+b), \frac{1}{2}(b-a), c$ | $\frac{1}{2},0,0; 0,\frac{1}{2},0$ | | | $-y,-x,z$ | $4 \cdot 1 \cdot 2$ |
| | | $a = b, \gamma = 120°$ | $cmmm$ | $\frac{1}{2}a, \frac{1}{2}a + b, c$ | | | | $x-y,-y,-z$ | $1 \cdot 1 \cdot 2$ |

| # | | | | | | | | | |
|---|---|---|---|---|---|---|---|---|---|
| 8 | p211 | General Metric | $p^1mmm$ | $\varepsilon_1 a, \frac{1}{2}b, c$ | $x_1,0,0; \frac{1}{2},0,0$ | $(0,0,0)$ | | $-x,y,-z$ | $(1\infty_\parallel)\cdot 2\cdot 2$ |
| | | $a=b$ | $p^1mmm$ | $\varepsilon_1 a, \frac{1}{2}b, c$ | $x_1,0,0; \frac{1}{2},0,0$ | $(0,0,0)$ | | $-x,y,-z$ | $(1\infty_\parallel)\cdot 2\cdot 2$ |
| 9 | $p2_1 11$ | General Metric | $p^1mmm$ | $\varepsilon_1 a, \frac{1}{2}b, c$ | $x_1,0,0; \frac{1}{2},0,0$ | $(0,0,0)$ | | $-x,y,-z$ | $(1\infty_\parallel)\cdot 2\cdot 2$ |
| | | $a=b$ | $p^1mmm$ | $\varepsilon_1 a, \frac{1}{2}b, c$ | $x_1,0,0; \frac{1}{2},0,0$ | $(0,0,0)$ | | $-x,y,-z$ | $(1\infty_\parallel)\cdot 2\cdot 2$ |
| 10 | $c211$ | General Metric | $p^1mmm$ | $\varepsilon_1 a, \frac{1}{2}b, c$ | $x_1,0,0; \frac{1}{2},0,0$ | $(0,0,0)$ | | $-x,y,-z$ | $(1\infty_\parallel)\cdot 2\cdot 2$ |
| | | $a=b$ | $p^1mmm$ | $\varepsilon_1 a, \frac{1}{2}b, c$ | $x_1,0,0; \frac{1}{2},0,0$ | $(0,0,0)$ | | $-x,y,-z$ | $(1\infty_\parallel)\cdot 2\cdot 2$ |
| 11 | $pm11$ | General Metric | $P^2mmm$ | $\frac{1}{2}a, \varepsilon_2 b, \varepsilon_3 c$ | $\frac{1}{2},0,0;\ 0,x_2,0;\ 0,0,x_3$ | $(0,0,0)$ | | $-x,y,-z$ | $(\infty_\perp \infty_\parallel)\cdot 2\cdot 2$ |
| | | $a=b$ | $P^2mmm$ | $\frac{1}{2}a, \varepsilon_2 b, \varepsilon_3 c$ | $\frac{1}{2},0,0;\ 0,x_2,0;\ 0,0,x_3$ | $(0,0,0)$ | | $-x,y,-z$ | $(\infty_\perp \infty_\parallel)\cdot 2\cdot 2$ |
| 12 | $pb11$ | General Metric | $P^2mmm$ | $\frac{1}{2}a, \varepsilon_2 b, \varepsilon_3 c$ | $\frac{1}{2},0,0;\ 0,x_2,0;\ 0,0,x_3$ | $(0,0,0)$ | | $-x,y,-z$ | $(\infty_\perp \infty_\parallel)\cdot 2\cdot 2$ |
| | | $a=b$ | $P^2mmm$ | $\frac{1}{2}a, \varepsilon_2 b, \varepsilon_3 c$ | $\frac{1}{2},0,0;\ 0,x_2,0;\ 0,0,x_3$ | $(0,0,0)$ | | $-x,y,-z$ | $(\infty_\perp \infty_\parallel)\cdot 2\cdot 2$ |
| 13 | $cm11$ | General Metric | $P^2mmm$ | $\frac{1}{2}a, \varepsilon_2 b, \varepsilon_3 c$ | $\frac{1}{2},0,0;\ 0,x_2,0;\ 0,0,x_3$ | $(0,0,0)$ | | $-x,y,-z$ | $(\infty_\perp \infty_\parallel)\cdot 2\cdot 2$ |
| | | $a=b$ | $P^2mmm$ | $\frac{1}{2}a, \varepsilon_2 b, \varepsilon_3 c$ | $\frac{1}{2},0,0;\ 0,x_2,0;\ 0,0,x_3$ | $(0,0,0)$ | | $-x,y,-z$ | $(\infty_\perp \infty_\parallel)\cdot 2\cdot 2$ |
| 14 | $p2/m11$ | General Metric | $pmmm$ | $\frac{1}{2}a, \frac{1}{2}b, c$ | $\frac{1}{2},0,0; 0,\frac{1}{2},0$ | | | $-x,y,-z$ | $4\cdot 1\cdot 2$ |
| | | $a=b$ | $pmmm$ | $\frac{1}{2}a, \frac{1}{2}b, c$ | $\frac{1}{2},0,0; 0,\frac{1}{2},0$ | | | $-x,y,-z$ | $4\cdot 1\cdot 2$ |
| 15 | $p2_1/m11$ | General Metric | $pmmm$ | $\frac{1}{2}a, \frac{1}{2}b, c$ | $\frac{1}{2},0,0; 0,\frac{1}{2},0$ | | | $-x,y,-z$ | $4\cdot 1\cdot 2$ |
| | | $a=b$ | $pmmm$ | $\frac{1}{2}a, \frac{1}{2}b, c$ | $\frac{1}{2},0,0; 0,\frac{1}{2},0$ | | | $-x,y,-z$ | $4\cdot 1\cdot 2$ |
| 16 | $p2/b11$ | General Metric | $pmmm$ | $\frac{1}{2}a, \frac{1}{2}b, c$ | $\frac{1}{2},0,0; 0,\frac{1}{2},0$ | | | $-x,y,-z$ | $4\cdot 1\cdot 2$ |
| | | $a=b$ | $pmmm$ | $\frac{1}{2}a, \frac{1}{2}b, c$ | $\frac{1}{2},0,0; 0,\frac{1}{2},0$ | | | $-x,y,-z$ | $4\cdot 1\cdot 2$ |
| 17 | $p2_1/b11$ | General Metric | $pmmm$ | $\frac{1}{2}a, \frac{1}{2}b, c$ | $\frac{1}{2},0,0; 0,\frac{1}{2},0$ | | | $-x,y,-z$ | $4\cdot 1\cdot 2$ |
| | | $a=b$ | $pmmm$ | $\frac{1}{2}a, \frac{1}{2}b, c$ | $\frac{1}{2},0,0; 0,\frac{1}{2},0$ | | | $-x,y,-z$ | $4\cdot 1\cdot 2$ |
| 18 | $c2/m11$ | General Metric | $pmmm$ | $\frac{1}{2}a, \frac{1}{2}b, c$ | $\frac{1}{2},0,0; 0,\frac{1}{2},0$ | | | $-x,y,-z$ | $4\cdot 1\cdot 2$ |
| | | $a=b$ | $pmmm$ | $\frac{1}{2}a, \frac{1}{2}b, c$ | $\frac{1}{2},0,0; 0,\frac{1}{2},0$ | | | $-x,y,-z$ | $4\cdot 1\cdot 2$ |
| 19 | $p222$ | General Metric | $pmmm$ | $\frac{1}{2}a, \frac{1}{2}b, c$ | $\frac{1}{2},0,0; 0,\frac{1}{2},0$ | $(0,0,0)$ | | | $4\cdot 2\cdot 1$ |
| | | $a=b$ | $p4/mmm$ | $\frac{1}{2}a, \frac{1}{2}b, c$ | $\frac{1}{2},0,0; 0,\frac{1}{2},0$ | $(0,0,0)$ | | $-y,x,z$ | $4\cdot 2\cdot 2$ |

| | | | | | | | | |
|---|---|---|---|---|---|---|---|---|
| 20 | $p2_122$ | General Metric | $pmmm$ | $\frac{1}{2}\boldsymbol{a}, \frac{1}{2}\boldsymbol{b}, \boldsymbol{c}$ | $\frac{1}{2},0,0; 0,\frac{1}{2},0$ | $(0,0,0)$ | | $4 \cdot 2 \cdot 1$ |
| | | $a = b$ | $pmmm$ | $\frac{1}{2}\boldsymbol{a}, \frac{1}{2}\boldsymbol{b}, \boldsymbol{c}$ | $\frac{1}{2},0,0; 0,\frac{1}{2},0$ | $(0,0,0)$ | | $4 \cdot 2 \cdot 1$ |
| 21 | $p2_12_12$ | General Metric | $pmmm$ | $\frac{1}{2}\boldsymbol{a}, \frac{1}{2}\boldsymbol{b}, \boldsymbol{c}$ | $\frac{1}{2},0,0; 0,\frac{1}{2},0$ | $(0,0,0)$ | | $4 \cdot 2 \cdot 1$ |
| | | $a = b$ | $p4/mmm$ | $\frac{1}{2}\boldsymbol{a}, \frac{1}{2}\boldsymbol{b}, \boldsymbol{c}$ | $\frac{1}{2},0,0; 0,\frac{1}{2},0$ | $(0,0,0)$ | $-y, x, z$ | $4 \cdot 2 \cdot 2$ |
| 22 | $c222$ | General Metric | $pmmm$ | $\frac{1}{2}\boldsymbol{a}, \frac{1}{2}\boldsymbol{b}, \boldsymbol{c}$ | $\frac{1}{2},0,0; 0,\frac{1}{2},0$ | $(0,0,0)$ | | $4 \cdot 2 \cdot 1$ |
| | | $a = b$ | $p4/mmm$ | $\frac{1}{2}\boldsymbol{a}, \frac{1}{2}\boldsymbol{b}, \boldsymbol{c}$ | $\frac{1}{2},0,0; 0,\frac{1}{2},0$ | $(0,0,0)$ | $-y, x, z$ | $4 \cdot 2 \cdot 2$ |
| 23 | $pmm2$ | General Metric | $P^1mmm$ | $\frac{1}{2}\boldsymbol{a}, \frac{1}{2}\boldsymbol{b}, \varepsilon_3\boldsymbol{c}$ | $\frac{1}{2},0,0; 0,\frac{1}{2},0; 0,0,x_3$ | $(0,0,0)$ | | $(\infty_\perp 4) \cdot 2 \cdot 1$ |
| | | $a = b$ | $P^14/mmm$ | $\frac{1}{2}\boldsymbol{a}, \frac{1}{2}\boldsymbol{b}, \varepsilon_3\boldsymbol{c}$ | $\frac{1}{2},0,0; 0,\frac{1}{2},0; 0,0,x_3$ | $(0,0,0)$ | $-y, x, z$ | $(\infty_\perp 4) \cdot 2 \cdot 2$ |
| 24 | $pma2$ | General Metric | $P^1mmm$ | $\frac{1}{2}\boldsymbol{a}, \frac{1}{2}\boldsymbol{b}, \varepsilon_3\boldsymbol{c}$ | $\frac{1}{2},0,0; 0,\frac{1}{2},0; 0,0,x_3$ | $(0,0,0)$ | | $(\infty_\perp 4) \cdot 2 \cdot 1$ |
| | | $a = b$ | $P^1mmm$ | $\frac{1}{2}\boldsymbol{a}, \frac{1}{2}\boldsymbol{b}, \varepsilon_3\boldsymbol{c}$ | $\frac{1}{2},0,0; 0,\frac{1}{2},0; 0,0,x_3$ | $(0,0,0)$ | | $(\infty_\perp 4) \cdot 2 \cdot 1$ |
| 25 | $pba2$ | General Metric | $P^1mmm$ | $\frac{1}{2}\boldsymbol{a}, \frac{1}{2}\boldsymbol{b}, \varepsilon_3\boldsymbol{c}$ | $\frac{1}{2},0,0; 0,\frac{1}{2},0; 0,0,x_3$ | $(0,0,0)$ | | $(\infty_\perp 4) \cdot 2 \cdot 1$ |
| | | $a = b$ | $P^14/mmm$ | $\frac{1}{2}\boldsymbol{a}, \frac{1}{2}\boldsymbol{b}, \varepsilon_3\boldsymbol{c}$ | $\frac{1}{2},0,0; 0,\frac{1}{2},0; 0,0,x_3$ | $(0,0,0)$ | $-y, x, z$ | $(\infty_\perp 4) \cdot 2 \cdot 2$ |
| 26 | $cmm2$ | General Metric | $P^1mmm$ | $\frac{1}{2}\boldsymbol{a}, \frac{1}{2}\boldsymbol{b}, \varepsilon_3\boldsymbol{c}$ | $\frac{1}{2},0,0; 0,\frac{1}{2},0; 0,0,x_3$ | $(0,0,0)$ | | $(\infty_\perp 4) \cdot 2 \cdot 1$ |
| | | $a = b$ | $P^14/mmm$ | $\frac{1}{2}\boldsymbol{a}, \frac{1}{2}\boldsymbol{b}, \varepsilon_3\boldsymbol{c}$ | $\frac{1}{2},0,0; 0,\frac{1}{2},0; 0,0,x_3$ | $(0,0,0)$ | $-y, x, z$ | $(\infty_\perp 4) \cdot 2 \cdot 2$ |
| 27 | $pm2m$ | General Metric | $p^1mmm$ | $\frac{1}{2}\boldsymbol{a}, \varepsilon_2\boldsymbol{b}, \boldsymbol{c}$ | $\frac{1}{2},0,0; 0,x_2,0$ | $(0,0,0)$ | | $(1\infty_\parallel) \cdot 2 \cdot 1$ |
| | | $a = b$ | $p^1mmm$ | $\frac{1}{2}\boldsymbol{a}, \varepsilon_2\boldsymbol{b}, \boldsymbol{c}$ | $\frac{1}{2},0,0; 0,x_2,0$ | $(0,0,0)$ | | $(1\infty_\parallel) \cdot 2 \cdot 1$ |
| 28 | $pm2_1b$ | General Metric | $p^1mmm$ | $\frac{1}{2}\boldsymbol{a}, \varepsilon_2\boldsymbol{b}, \boldsymbol{c}$ | $\frac{1}{2},0,0; 0,x_2,0$ | $(0,0,0)$ | | $(1\infty_\parallel) \cdot 2 \cdot 1$ |
| | | $a = b$ | $p^1mmm$ | $\frac{1}{2}\boldsymbol{a}, \varepsilon_2\boldsymbol{b}, \boldsymbol{c}$ | $\frac{1}{2},0,0; 0,x_2,0$ | $(0,0,0)$ | | $(1\infty_\parallel) \cdot 2 \cdot 1$ |
| 29 | $pb2_1b$ | General Metric | $p^1mmm$ | $\frac{1}{2}\boldsymbol{a}, \varepsilon_2\boldsymbol{b}, \boldsymbol{c}$ | $\frac{1}{2},0,0; 0,x_2,0$ | $(0,0,0)$ | | $(1\infty_\parallel) \cdot 2 \cdot 1$ |
| | | $a = b$ | $p^1mmm$ | $\frac{1}{2}\boldsymbol{a}, \varepsilon_2\boldsymbol{b}, \boldsymbol{c}$ | $\frac{1}{2},0,0; 0,x_2,0$ | $(0,0,0)$ | | $(1\infty_\parallel) \cdot 2 \cdot 1$ |
| 30 | $pb2b$ | General Metric | $p^1mmm$ | $\frac{1}{2}\boldsymbol{a}, \varepsilon_2\boldsymbol{b}, \boldsymbol{c}$ | $\frac{1}{2},0,0; 0,x_2,0$ | $(0,0,0)$ | | $(1\infty_\parallel) \cdot 2 \cdot 1$ |
| | | $a = b$ | $p^1mmm$ | $\frac{1}{2}\boldsymbol{a}, \varepsilon_2\boldsymbol{b}, \boldsymbol{c}$ | $\frac{1}{2},0,0; 0,x_2,0$ | $(0,0,0)$ | | $(1\infty_\parallel) \cdot 2 \cdot 1$ |
| 31 | $pm2a$ | General Metric | $p^1mmm$ | $\frac{1}{2}\boldsymbol{a}, \varepsilon_2\boldsymbol{b}, \boldsymbol{c}$ | $\frac{1}{2},0,0; 0,x_2,0$ | $(0,0,0)$ | | $(1\infty_\parallel) \cdot 2 \cdot 1$ |
| | | $a = b$ | $p^1mmm$ | $\frac{1}{2}\boldsymbol{a}, \varepsilon_2\boldsymbol{b}, \boldsymbol{c}$ | $\frac{1}{2},0,0; 0,x_2,0$ | $(0,0,0)$ | | $(1\infty_\parallel) \cdot 2 \cdot 1$ |
| 32 | $pm2_1n$ | General Metric | $p^1mmm$ | $\frac{1}{2}a, \varepsilon_2 b, c$ | $\frac{1}{2},0,0; 0,x_2,0$ | $(0,0,0)$ | | $(1\infty_\parallel)\cdot 2\cdot 1$ |
| | | $a=b$ | $p^1mmm$ | $\frac{1}{2}a, \varepsilon_2 b, c$ | $\frac{1}{2},0,0; 0,x_2,0$ | $(0,0,0)$ | | $(1\infty_\parallel)\cdot 2\cdot 1$ |
| 33 | $pb2_1a$ | General Metric | $p^1mmm$ | $\frac{1}{2}a, \varepsilon_2 b, c$ | $\frac{1}{2},0,0; 0,x_2,0$ | $(0,0,0)$ | | $(1\infty_\parallel)\cdot 2\cdot 1$ |
| | | $a=b$ | $p^1mmm$ | $\frac{1}{2}a, \varepsilon_2 b, c$ | $\frac{1}{2},0,0; 0,x_2,0$ | $(0,0,0)$ | | $(1\infty_\parallel)\cdot 2\cdot 1$ |
| 34 | $pb2n$ | General Metric | $p^1mmm$ | $\frac{1}{2}a, \varepsilon_2 b, c$ | $\frac{1}{2},0,0; 0,x_2,0$ | $(0,0,0)$ | | $(1\infty_\parallel)\cdot 2\cdot 1$ |
| | | $a=b$ | $p^1mmm$ | $\frac{1}{2}a, \varepsilon_2 b, c$ | $\frac{1}{2},0,0; 0,x_2,0$ | $(0,0,0)$ | | $(1\infty_\parallel)\cdot 2\cdot 1$ |
| 35 | $cm2m$ | General Metric | $p^1mmm$ | $\frac{1}{2}a, \varepsilon_2 b, c$ | $\frac{1}{2},0,0; 0,x_2,0$ | $(0,0,0)$ | | $(1\infty_\parallel)\cdot 2\cdot 1$ |
| | | $a=b$ | $p^1mmm$ | $\frac{1}{2}a, \varepsilon_2 b, c$ | $\frac{1}{2},0,0; 0,x_2,0$ | $(0,0,0)$ | | $(1\infty_\parallel)\cdot 2\cdot 1$ |
| 36 | $cm2e$ | General Metric | $p^1mmm$ | $\frac{1}{2}a, \varepsilon_2 b, c$ | $\frac{1}{2},0,0; 0,x_2,0$ | $(0,0,0)$ | | $(1\infty_\parallel)\cdot 2\cdot 1$ |
| | | $a=b$ | $p^1mmm$ | $\frac{1}{2}a, \varepsilon_2 b, c$ | $\frac{1}{2},0,0; 0,x_2,0$ | $(0,0,0)$ | | $(1\infty_\parallel)\cdot 2\cdot 1$ |
| 37 | $pmmm$ | General Metric | $pmmm$ | $\frac{1}{2}a, \frac{1}{2}b, c$ | $\frac{1}{2},0,0; 0,\frac{1}{2},0$ | | | $4\cdot 1\cdot 1$ |
| | | $a=b$ | $p4/mmm$ | $\frac{1}{2}a, \frac{1}{2}b, c$ | $\frac{1}{2},0,0; 0,\frac{1}{2},0$ | | $-y,x,z$ | $4\cdot 1\cdot 2$ |
| 38 | $pmaa$ | General Metric | $pmmm$ | $\frac{1}{2}a, \frac{1}{2}b, c$ | $\frac{1}{2},0,0; 0,\frac{1}{2},0$ | | | $4\cdot 1\cdot 1$ |
| | | $a=b$ | $pmmm$ | $\frac{1}{2}a, \frac{1}{2}b, c$ | $\frac{1}{2},0,0; 0,\frac{1}{2},0$ | | | $4\cdot 1\cdot 1$ |
| 39 | $pban$ | General Metric | $pmmm$ | $\frac{1}{2}a, \frac{1}{2}b, c$ | $\frac{1}{2},0,0; 0,\frac{1}{2},0$ | | | $4\cdot 1\cdot 1$ |
| | | $a=b$ | $p4/mmm$ | $\frac{1}{2}a, \frac{1}{2}b, c$ | $\frac{1}{2},0,0; 0,\frac{1}{2},0$ | | $-y,x,z$ | $4\cdot 1\cdot 2$ |
| 40 | $pmam$ | General Metric | $pmmm$ | $\frac{1}{2}a, \frac{1}{2}b, c$ | $\frac{1}{2},0,0; 0,\frac{1}{2},0$ | | | $4\cdot 1\cdot 1$ |
| | | $a=b$ | $pmmm$ | $\frac{1}{2}a, \frac{1}{2}b, c$ | $\frac{1}{2},0,0; 0,\frac{1}{2},0$ | | | $4\cdot 1\cdot 1$ |
| 41 | $pmma$ | General Metric | $pmmm$ | $\frac{1}{2}a, \frac{1}{2}b, c$ | $\frac{1}{2},0,0; 0,\frac{1}{2},0$ | | | $4\cdot 1\cdot 1$ |
| | | $a=b$ | $pmmm$ | $\frac{1}{2}a, \frac{1}{2}b, c$ | $\frac{1}{2},0,0; 0,\frac{1}{2},0$ | | | $4\cdot 1\cdot 1$ |
| 42 | $pman$ | General Metric | $pmmm$ | $\frac{1}{2}a, \frac{1}{2}b, c$ | $\frac{1}{2},0,0; 0,\frac{1}{2},0$ | | | $4\cdot 1\cdot 1$ |
| | | $a=b$ | $pmmm$ | $\frac{1}{2}a, \frac{1}{2}b, c$ | $\frac{1}{2},0,0; 0,\frac{1}{2},0$ | | | $4\cdot 1\cdot 1$ |
| 43 | $pbaa$ | General Metric | $pmmm$ | $\frac{1}{2}a, \frac{1}{2}b, c$ | $\frac{1}{2},0,0; 0,\frac{1}{2},0$ | | | $4\cdot 1\cdot 1$ |
| | | $a=b$ | $pmmm$ | $\frac{1}{2}a, \frac{1}{2}b, c$ | $\frac{1}{2},0,0; 0,\frac{1}{2},0$ | | | $4\cdot 1\cdot 1$ |
| 44 | $p\bar{b}am$ | General Metric | $pmmm$ | $\frac{1}{2}a, \frac{1}{2}b, c$ | $\frac{1}{2},0,0; 0,\frac{1}{2},0$ | | | $4 \cdot 1 \cdot 1$ |
| | | $a = b$ | $p4/mmm$ | $\frac{1}{2}a, \frac{1}{2}b, c$ | $\frac{1}{2},0,0; 0,\frac{1}{2},0$ | | $-y, x, z$ | $4 \cdot 1 \cdot 2$ |
| 45 | $p\bar{b}ma$ | General Metric | $pmmm$ | $\frac{1}{2}a, \frac{1}{2}b, c$ | $\frac{1}{2},0,0; 0,\frac{1}{2},0$ | | | $4 \cdot 1 \cdot 1$ |
| | | $a = b$ | $pmmm$ | $\frac{1}{2}a, \frac{1}{2}b, c$ | $\frac{1}{2},0,0; 0,\frac{1}{2},0$ | | | $4 \cdot 1 \cdot 1$ |
| 46 | $pmmn$ | General Metric | $pmmm$ | $\frac{1}{2}a, \frac{1}{2}b, c$ | $\frac{1}{2},0,0; 0,\frac{1}{2},0$ | | | $4 \cdot 1 \cdot 1$ |
| | | $a = b$ | $p4/mmm$ | $\frac{1}{2}a, \frac{1}{2}b, c$ | $\frac{1}{2},0,0; 0,\frac{1}{2},0$ | | $-y, x, z$ | $4 \cdot 1 \cdot 2$ |
| 47 | $cmmm$ | General Metric | $pmmm$ | $\frac{1}{2}a, \frac{1}{2}b, c$ | $\frac{1}{2},0,0; 0,\frac{1}{2},0$ | | | $2 \cdot 1 \cdot 1$ |
| | | $a = b$ | $p4/mmm$ | $\frac{1}{2}a, \frac{1}{2}b, c$ | $\frac{1}{2},0,0; 0,\frac{1}{2},0$ | | $-y, x, z$ | $2 \cdot 1 \cdot 2$ |
| 48 | $cmme$ | General Metric | $pmmm$ | $\frac{1}{2}a, \frac{1}{2}b, c$ | $\frac{1}{2},0,0; 0,\frac{1}{2},0$ | | | $2 \cdot 1 \cdot 1$ |
| | | $a = b$ | $p4/mmm$ | $\frac{1}{2}a, \frac{1}{2}b, c$ | $\frac{1}{2},0,0; 0,\frac{1}{2},0$ | | $-y+\frac{1}{4}, x+\frac{1}{4}, z$ | $2 \cdot 1 \cdot 2$ |
| 49 | $p4$ | | $P^14/mmm$ | $\frac{1}{2}(a+b), \frac{1}{2}(b-a), \varepsilon_3 c$ | $0,0,x_3; \frac{1}{2},\frac{1}{2},0$ | $(0,0,0)$ | $-x, y, -z$ | $(\infty_\perp 2) \cdot 2 \cdot 2$ |
| 50 | $p\bar{4}$ | | $p4/mmm$ | $\frac{1}{2}(a+b), \frac{1}{2}(b-a), c$ | $\frac{1}{2},\frac{1}{2},0$ | $(0,0,0)$ | $-x, y, -z$ | $2 \cdot 2 \cdot 2$ |
| 51 | $p4/m$ | | $p4/mmm$ | $\frac{1}{2}(a+b), \frac{1}{2}(b-a), c$ | $\frac{1}{2},\frac{1}{2},0$ | | $-x, y, -z$ | $2 \cdot 1 \cdot 2$ |
| 52 | $p4/n$ | | $p4/mmm$ $(mmm)$ | $\frac{1}{2}(a+b), \frac{1}{2}(b-a), c$ | $\frac{1}{2},\frac{1}{2},0$ | | $-x, y, -z$ | $2 \cdot 1 \cdot 2$ |
| 53 | $p422$ | | $p4/mmm$ | $\frac{1}{2}(a+b), \frac{1}{2}(b-a), c$ | $\frac{1}{2},\frac{1}{2},0$ | $(0,0,0)$ | | $2 \cdot 2 \cdot 1$ |
| 54 | $p42_12$ | | $p4/mmm$ | $\frac{1}{2}(a+b), \frac{1}{2}(b-a), c$ | $\frac{1}{2},\frac{1}{2},0$ | $(0,0,0)$ | | $2 \cdot 2 \cdot 1$ |
| 55 | $p4mm$ | | $P^14/mmm$ | $\frac{1}{2}(a+b), \frac{1}{2}(b-a), \varepsilon_3 c$ | $0,0,x_3; \frac{1}{2},\frac{1}{2},0$ | $(0,0,0)$ | | $(\infty_\perp 2) \cdot 2 \cdot 1$ |
| 56 | $p4bm$ | | $P^14/mmm$ | $\frac{1}{2}(a+b), \frac{1}{2}(b-a), \varepsilon_3 c$ | $0,0,x_3; \frac{1}{2},\frac{1}{2},0$ | $(0,0,0)$ | | $(\infty_\perp 2) \cdot 2 \cdot 1$ |
| 57 | $p\bar{4}2m$ | | $p4/mmm$ | $\frac{1}{2}(a+b), \frac{1}{2}(b-a), c$ | $\frac{1}{2},\frac{1}{2},0$ | $(0,0,0)$ | | $2 \cdot 2 \cdot 1$ |
| 58 | $p\bar{4}2_1m$ | | $p4/mmm$ | $\frac{1}{2}(a+b), \frac{1}{2}(b-a), c$ | $\frac{1}{2},\frac{1}{2},0$ | $(0,0,0)$ | | $2 \cdot 2 \cdot 1$ |
| 59 | $p\bar{4}m2$ | | $p4/mmm$ | $\frac{1}{2}(a+b), \frac{1}{2}(b-a), c$ | $\frac{1}{2},\frac{1}{2},0$ | $(0,0,0)$ | | $2 \cdot 2 \cdot 1$ |
| 60 | $p\bar{4}b2$ | | $p4/mmm$ | $\frac{1}{2}(a+b), \frac{1}{2}(b-a), c$ | $\frac{1}{2},\frac{1}{2},0$ | $(0,0,0)$ | | $2 \cdot 2 \cdot 1$ |

| # | | | | | | | | |
|---|---|---|---|---|---|---|---|---|
| 61 | $p4/mmm$ | | $p4/mmm$ | $\frac{1}{2}(a+b), \frac{1}{2}(b-a), c$ | $\frac{1}{2}, \frac{1}{2}, 0$ | | | $2 \cdot 1 \cdot 1$ |
| 62 | $p4/nbm$ | | $p4/mmm$ $(mmm)$ | $\frac{1}{2}(a+b), \frac{1}{2}(b-a), c$ | $\frac{1}{2}, \frac{1}{2}, 0$ | | | $2 \cdot 1 \cdot 1$ |
| 63 | $p4/mbm$ | | $p4/mmm$ | $\frac{1}{2}(a+b), \frac{1}{2}(b-a), c$ | $\frac{1}{2}, \frac{1}{2}, 0$ | | | $2 \cdot 1 \cdot 1$ |
| 64 | $p4/nmm$ | | $p4/mmm$ $(mmm)$ | $\frac{1}{2}(a+b), \frac{1}{2}(b-a), c$ | $\frac{1}{2}, \frac{1}{2}, 0$ | | | $2 \cdot 1 \cdot 1$ |
| 65 | $p3$ | | $P^1 6/mmm$ | $\frac{1}{3}(2a-b), \frac{1}{3}(b-a), \epsilon_3 c$ | $\frac{2}{3}, \frac{1}{3}, 0$ | $(0,0,0)$ | $-x,-y,z; y,x,-z$ | $(\infty_\perp 3) \cdot 2 \cdot 4$ |
| 66 | $p\bar{3}$ | | $p6/mmm$ | $a, b, c$ | | | $-x,-y,z; y,x,-z$ | $1 \cdot 1 \cdot 4$ |
| 67 | $p312$ | | $p6/mmm$ | $\frac{1}{3}(2a-b), \frac{1}{3}(b-a), c$ | $\frac{2}{3}, \frac{1}{3}, 0$ | $(0,0,0)$ | $-x,-y,z$ | $3 \cdot 2 \cdot 2$ |
| 68 | $p321$ | | $p6/mmm$ | $a, b, c$ | | $(0,0,0)$ | $-x,-y,z$ | $1 \cdot 2 \cdot 2$ |
| 69 | $p3m1$ | | $P^1 6/mmm$ | $\frac{1}{3}(2a-b), \frac{1}{3}(b-a), \epsilon_3 c$ | $0,0, x_3; \frac{2}{3}, \frac{1}{3}, 0$ | $(0,0,0)$ | $-x,-y,z$ | $(\infty_\perp 3) \cdot 2 \cdot 2$ |
| 70 | $p31m$ | | $P^1 6/mmm$ | $a, b, \epsilon_3 c$ | $0,0, x_3$ | $(0,0,0)$ | $-x,-y,z$ | $(\infty_\perp 1) \cdot 2 \cdot 2$ |
| 71 | $p\bar{3}1m$ | | $p6/mmm$ | $a, b, c$ | | | $-x,-y,z$ | $1 \cdot 1 \cdot 2$ |
| 72 | $p\bar{3}m1$ | | $p6/mmm$ | $a, b, c$ | | | $-x,-y,z$ | $1 \cdot 1 \cdot 2$ |
| 73 | $p6$ | | $P^1 6/mmm$ | $a, b, \epsilon_3 c$ | $0,0, x_3$ | $(0,0,0)$ | $y,x,-z$ | $(\infty_\perp 1) \cdot 2 \cdot 2$ |
| 74 | $p\bar{6}$ | | $p6/mmm$ | $\frac{1}{3}(2a-b), \frac{1}{3}(b-a), c$ | $\frac{2}{3}, \frac{1}{3}, 0$ | $(0,0,0)$ | $y,x,-z$ | $3 \cdot 2 \cdot 2$ |
| 75 | $p6/m$ | | $p6/mmm$ | $a, b, c$ | | | $y,x,-z$ | $1 \cdot 1 \cdot 2$ |
| 76 | $p622$ | | $p6/mmm$ | $a, b, c$ | | $(0,0,0)$ | | $1 \cdot 2 \cdot 1$ |
| 77 | $p6mm$ | | $P^1 6/mmm$ | $a, b, \epsilon_3 c$ | $0,0, x_3$ | $(0,0,0)$ | | $(\infty_\perp 1) \cdot 2 \cdot 1$ |
| 78 | $p\bar{6}m2$ | | $p6/mmm$ | $\frac{1}{3}(2a-b), \frac{1}{3}(b-a), c$ | $\frac{2}{3}, \frac{1}{3}, 0$ | $(0,0,0)$ | | $3 \cdot 2 \cdot 1$ |
| 79 | $p\bar{6}2m$ | | $p6/mmm$ | $a, b, c$ | | $(0,0,0)$ | | $1 \cdot 2 \cdot 1$ |
| 80 | $p6/mmm$ | | $p6/mmm$ | $a, b, c$ | | | | $1 \cdot 1 \cdot 1$ |